\documentclass[twocolumn,aps,showpacs,prx,amsmath,amssymb,floatfix,superscriptaddress]{revtex4-1}
\usepackage{color}
\usepackage{graphicx}% Include figure files
\usepackage{dcolumn}% Align table columns on decimal point
\usepackage{bm}% bold math
\usepackage{array}
\usepackage{float}
\usepackage{supertabular}
\usepackage{longtable}
\usepackage{mathrsfs}
\usepackage{txfonts}
\usepackage{wasysym}
\definecolor{db}{rgb}{0,0,0.4}
\usepackage[usenames,dvipsnames]{xcolor}
\begin{document}

\title{
Interplay of Anderson localization and quench dynamics}

 \author{Armin Rahmani}
\affiliation{Department of Physics and Astronomy, Western Washington University, Bellingham, Washington 98225, USA}

\author{Smitha Vishveshwara}
\affiliation{
Department of Physics, University of Illinois at Urbana-Champaign, Urbana, Illinois 61801, USA}

\date{\today}

\begin{abstract}
In the context of an isolated three-dimensional noninteracting fermionic lattice system, we study the effects of a sudden quantum quench between a disorder-free situation and one in which disorder results in a mobility edge and associated Anderson localization.  Salient post-quench features  hinge upon the overlap between momentum states and post-quench eigenstates and whether these latter states are extended or localized. We find that the post-quench momentum distribution directly reflects these overlaps. For the local density, we show that disorder generically prevents the equilibration of quantum expectation values to a steady state and that the persistent fluctuations have a nonmonotonic dependence on the strength of disorder. We identify two distinct types of fluctuations, namely, temporal fluctuations describing the time-dependent fluctuations of the local density around its time average and sample-to-sample fluctuations characterizing the variations of these time averages from one realization of disorder to another. We demonstrate that both of these fluctuations vanish for extremely extended as well as extremely localized states, peaking at some intermediate value.

\end{abstract}
 
\maketitle

\section{INTRODUCTION}

\begin{figure}
\begin{center}
 \includegraphics[width =\columnwidth]{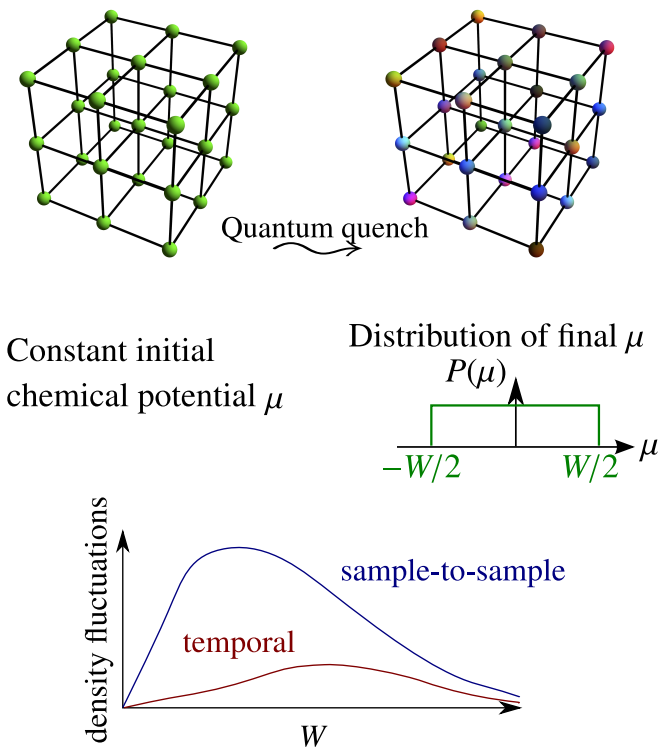}
 \caption{Top: Schematic of the quench protocol. The system is initially in the ground state of a clean tight-binding model of spinless fermions at constant chemical potential $\mu=0$. A disorder potential with uniform distribution and strength $W$ (represented by the color of the lattice sites) is suddenly turned on at $t=0$. Bottom: The nonmonotonic behavior of the post-quench density fluctuations with the strength of disorder.}
 \label{fig:1}
 \end{center}
\end{figure}

Generically, many-body quantum systems have two robust distinct fates after a quantum quench: thermalization and localization (see Ref.~\cite{Nandkishore2015} and the references therein). In the former case, the system effectively serves as a heat bath for small enough subsystems, resulting in equilibration to a steady state and the distribution of excess energy (which is deposited into the system after the quench) in an almost thermal manner. This behavior stems from the so called eigenstate thermalization hypothesis~\cite{Deutsch1991,Srednicki1994,Tasaki1998,Rigol2008}. In the latter case, as shown in the seminal work of Anderson~\cite{Anderson1958}, the flow of energy is restricted due to the presence of disorder, and systems can not act as reservoirs for themselves. The phenomenology of these nonthermalizing systems includes two categories: (i) many-body localization, where interactions play an important role~\cite{Basko2006, Oganesyan2007, Pal2010} and (ii) the simpler case of single-particle Anderson localization, where interactions are either absent or unimportant. Signatures of quantum quench dynamics for both paradigms of thermalization and localization are of great interest~\cite{Polkovnikov2011, Greiner2002, Kinoshita2006, Calabrese2006,Kollath2007,Cramer2008,Yukalov2011,Eisert2014,Canovi2011,Bardarson2012,Chandran2013,Zangara2013,Vosk2013,Sorg2014,Tang2015}.

Historically, the bulk of the studies in the localization literature have focused on the single-particle case. Despite the absence of interactions, the physics of Anderson localization is very rich.  In recent years, there has been a surge of interest in many-body localization, where interactions give rise to even richer phenomena. Substantial progress in understanding the nature of the many-body localization has been made by studying the novel question of the \textit{interplay of disorder and quench dynamics}~\cite{Canovi2011,Bardarson2012,Zangara2013,Vosk2013,Tang2015}.
Surprisingly, however, despite the large body of work on single-particle localization, this particular aspect, namely, the effects of disorder alone on the quench dynamics, has remained relatively unexplored in the literature. Only recently, a few studies have begun to  address this problem. In one spatial dimension, where all single-particle states are localized even for infinitesimally small disorder, it was shown that localization can prevent the emergence of a steady state~\cite{Ziraldo2012,Ziraldo2013}. There has also been related studies on the effects of quantum dynamics in systems with quasiperiodic potentials in one-dimensional models, where either all states are extended or localized depending on the strength of the quasiperiodic potential~\cite{Gramsch2012,Ribeiro2013,He2013,Wang2017}. (Although quasi-periodic potentials do not represent an ensemble of disorder realizations, they are expected to capture some aspects of the pertinent physics. Other aspects such as sample-to-sample fluctuations rely on having a true disorder ensemble.)

Here we study quench dynamics in the canonical three-dimensional fermionic model exhibiting the Anderson localization transition.  Our thrust lies in demonstrating that Anderson localization physics and associated features have a direct effect on the post-quench dynamics, particularly on fluctuation properties of observables. 
The post-quench behavior of observables, including their fluctuations, is intimately related to the nature and the statistics of the disordered wave functions  (see Ref.~\cite{Mirlin2000} for a review). In the absence of disorder, these wave functions are plane waves. At infinite disorder, all disorder realizations give rise to the same set of on-site localized eigenfunctions. It is for intermediate strength of disorder that the salient features of the localization transition and crossover between these two limits appear. Different realizations of disorder give rise to an ensemble of quantum eigenstates. Some states are extended and others are  localized. A change between these two different behaviors involves a phase transition and an associated diverging localization length.  The two phases are separated by a mobility edge.

Even within the localization regime at higher (but finite) disorder, there is a wide distribution of localization lengths.

To probe the consequences of these features on dynamics, we focus on quantum quenches where the system starts in the ground state of a clean three-dimensional tight-binding Hamiltonian of spinless fermions with translationally invariant nearest-neighbor hopping at fixed particle number. As shown in Fig.~\ref{fig:1}, a disordered chemical potential is then suddenly turned on. The post-quench time evolution is governed by the nature of the single-particle eigenfunctions of the final disordered Hamiltonian (how localized they are, what is the distribution of the localization lengths,  etc.) and in particular their overlaps with the plane-wave eigenfunctions of the initial clean system. The more localized the final wave functions are, the more uniform these overlaps become. The final momentum distribution (averaged over time and disorder), which, through time-of-flight measurements,  is the most easily accessible observable in cold-atom experiments (see Ref.~\cite{Bloch2008} and the  references therein) captures this feature. It monotonically crosses over from a typical Fermi-Dirac step function for quenches to small disorder to a uniform distribution for quenches to large disorder.

Our main results concern the problem of the fluctuations of real-space density in our system. With regards to feasibility of measurement, once again, in cold atomic systems, new \textit{in situ} imaging techniques provide direct experimental access to the real-space density, giving a complementary picture to the time-of-flight measurements~\cite{Esteve2006,Gemelke2009,Muller2010,Bakr2010,Sherson2010}. Fluctuation phenomena are of particular interest  in disordered systems as they stem from multiple sources. As argued in Ref. ~\cite{Ziraldo2012}, the presence of disorder in the final Hamiltonian can prevent the relaxation of the system to a steady state in the type of systems considered here, resulting in \textit{persistent temporal fluctuations of various observables}. Here we perform a quantitative analysis of the temporal fluctuations of local density and show that they have a \textit{nonmonotnic} dependence on the strength of disorder~\footnote{Nonmonotonic dependence on the strength of disorder has also been observed in the noise magnitude of equilibrium disordered systems~\cite{Cohen1992}} .
% At very high strength of disorder, these fluctuations begin to decrease due to the competing effects of disorder. This leads to observable equilibration in the limit of infinite disorder. 

 Having an ensemble of final Hamiltonians (and consequently an ensemble of quenches) is one of the distinctive attributes of disordered systems. Most studies of quantum quenches, in which a parameter in the Hamiltonian changes for a system initially in the ground state, are described by a unique time-dependent wave function. In the quantum quench we consider here, the parameter undergoing the quench is a property of a distribution. Conceptually, this quench can be regarded as an ensemble of quantum quenches~\cite{Rahmani2013}: for each realization of disorder, the chemical potentials $\mu^{\bf x}_W$ are suddenly turned on and the system undergoes unitary evolution. Observables of interest are then averaged over the realizations of disorder. In addition to understanding the time evolution for individual realizations of disorder, the variations of the dynamics from one sample to another are therefore important in the full description of the quench. We thus consider a second type of fluctuations, namely, the fluctuations of the time averages of the local density from sample to sample. We find that these \textit{sample-to-sample fluctuations} also exhibit a nonmonotonic dependence on the strength of disorder.

Hence, as shown in  Fig.~\ref{fig:1}, both temporal and sample-to-sample density fluctuations on a given site  have a nonmonotonic dependence on the strength of disorder, peaking at intermediate values of disorder. Temporal fluctuations capture the absence of equilibration and persist for stronger disorder than the fluctuations between samples; the former peaks at stronger disorder than the latter.  A distinguishing feature of these fluctuations is that they appear to survive in the thermodynamic limit. Although our scaling analysis is done for small systems, we do not observe strong system-size dependence for sample-to-sample fluctuations. The temporal fluctuations do decrease with system size. Extrapolation is suggestive of the survival of these fluctuations in the thermodynamic limit. In the one-dimensional case, where numerical studies of much larger systems is possible, more compelling evidence for the survival of the temporal fluctuation in the thermodynamic limit was found in Ref. ~\cite{Ziraldo2012}. In equilibrium mesoscopic systems, many types of fluctuations generically vanish in the thermodynamic limit due to self-averaging. A classic example, where fluctuations are \textit{not} suppressed by self-averaging is conductance fluctuations~\cite{Altshuler1985,Lee1985}.

The outline of this paper is as follows. In Sec.~\ref{sec:2}, we present the model and discuss some of its important features. In  Sec.~\ref{sec:3}, we  focus on the behavior of wave function overlaps and of the momentum distribution. In Sec.~\ref{sec:4}, we present our results on the temporal and sample-to-sample fluctuations of density. We close the paper in Sec.~\ref{sec:5} with a brief summary and conclusions.

\section{MODEL AND THE QUANTUM QUENCH}	\label{sec:2}

In this work, we study the prototypical Anderson model of localization in three spatial dimensions, which exhibits a localization-induced metal-insulator transition.  The Hamiltonian describing the system is given by
\begin{equation}\label{eq:hamil}
H_W=-\Gamma\sum_{\langle {\bf x}{\bf y}\rangle}\left(c^\dagger_{\bf x} c_{\bf y}+c^\dagger_{\bf y} c_{\bf x}\right)+\sum_{\bf x} \mu^{\bf x}_W c^\dagger_{\bf x} c_{\bf x},
\end{equation}
where $c_{\bf x}$ is the fermionic annihilation operator on site ${\bf x}$ of a three-dimensional (3D) cubic lattice and $\langle {\bf x}{\bf y}\rangle$ indicates nearest-neighbor sites ${\bf x}$ and ${\bf y}$. The quantity $\mu^{\bf x}_W$ represents a random chemical potential drawn from a uniform distribution $\left[-{W\over 2},+{W\over 2}\right]$, with $W$ representing the strength of disorder. Hereafter we set the hopping amplitude to unity, $\Gamma=1$. We assume the system is an $L\times L\times L$ cubic lattice having periodic boundary conditions and that $M=L^3$ is the total number of lattice sites. As the total number of particles $N=\sum_{\bf x}  c^\dagger_{\bf x} c_{\bf x}$ is conserved, we study the dynamics in sectors with constant density $N/M$.

\begin{figure}
\begin{center}
 \includegraphics[width =0.64\columnwidth]{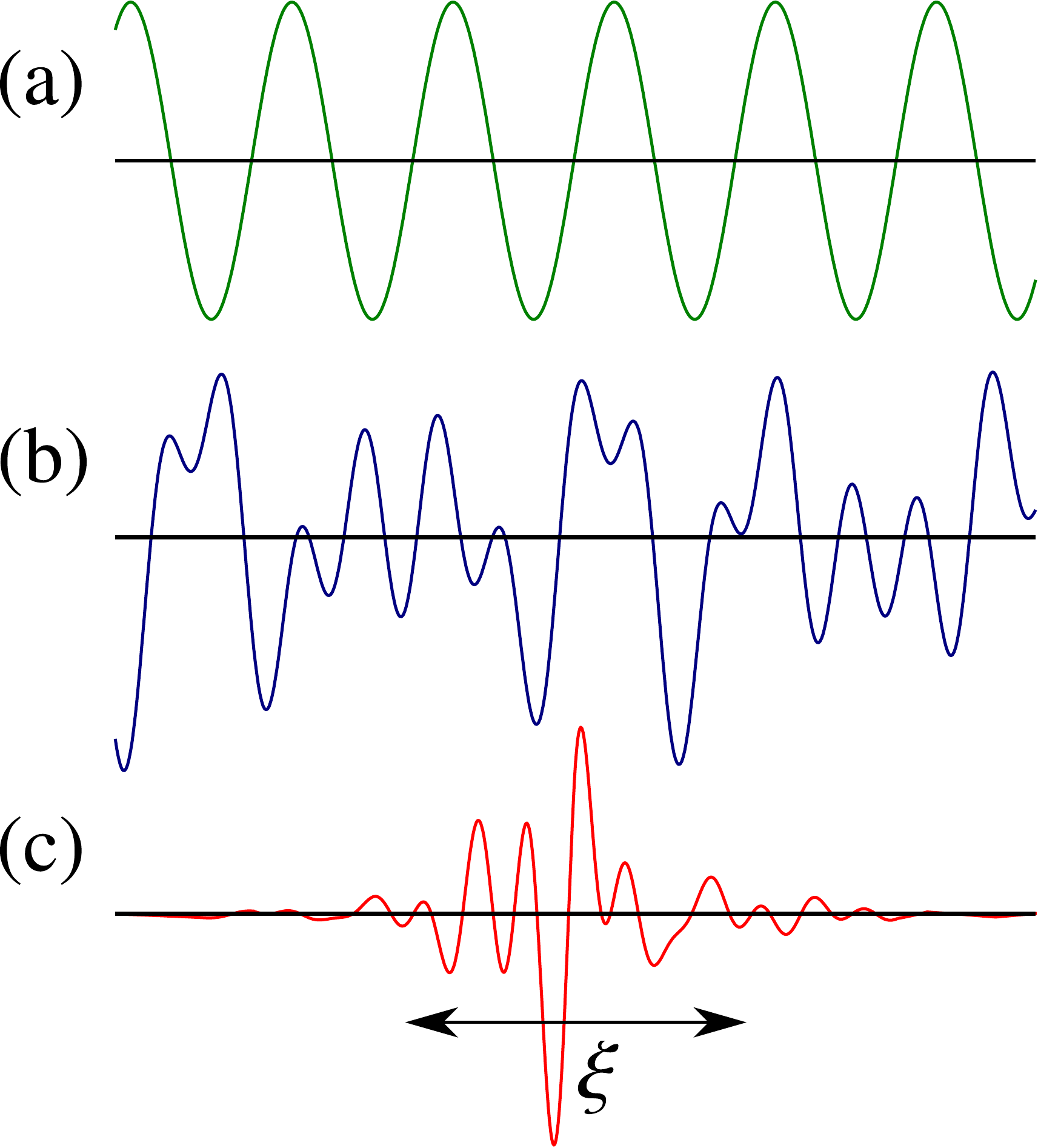}
 \caption{Real-space schematic of (a) a plane-wave state, (b) an extended disordered wave function, and (c) and a localized wave function.\label{fig:2}}
 \end{center}
\end{figure}

In the clean case ($W=0$), the Hamiltonian has translation invariance and momentum is a good quantum number. We can then write $H_0=\sum_{\bf k} \epsilon_{\bf k}c^\dagger_{\bf k}c_{\bf k}$, with dispersion relation $\epsilon_{\bf k}=-2\left(\cos k_x +\cos k_y +\cos k_z\right)$, where ${\bf k}=(k_x, k_y, k_z)$ is the momentum wave-vector. The single-particle wave functions of the clean system are plane waves, as depicted in Fig.~\ref{fig:2}(a) by a one-dimensional schematic:
 \begin{equation}\label{eq:pwave}
\psi^n_0({ \bf x})\equiv\langle { \bf x}|\psi^n_0\rangle={1\over \sqrt{M}}e^{i{\bf k}_n. { \bf x}},
\end{equation}
where $n$ is an integer that labels the momenta $ { \bf k}$ in the order of ascending energy $\epsilon_ { \bf k}$. Degenerate levels are arranged in an arbitrary manner. However, we always choose the number of fermions in such a way that all degenerate levels at a given energy are either empty or occupied so the arbitrary choice of the labeling is immaterial for physical properties.

 There is no gap in the single-particle spectrum for $W=0$ and the system is a Fermi-liquid metal at any density. When we add disorder to the system, the wave functions either (i) remain extended but acquire a characteristic mean free path as shown in Fig.~\ref{fig:2}(b) (roughly speaking for weak disorder the plane waves with momentum $\bf k$ are perturbed predominantly mixing with other plane waves of similar energy $\epsilon_{\bf k}$) or (ii) become localized  as shown in Fig.~\ref{fig:2}(c), where the wave function effectively has support in a region of characteristic length $\xi$ known as the localization length with an exponentially decaying envelope from a localization center (naturally this requires the mixing of many plane waves). At a critical energy $E_c$, which demarcates the boundary between localized and extended states, namely the mobility edge, the localization length on the localized side diverges as $|E-E_c|^{-\nu}$.  

\begin{figure}
\begin{center}
 \includegraphics[width =\columnwidth]{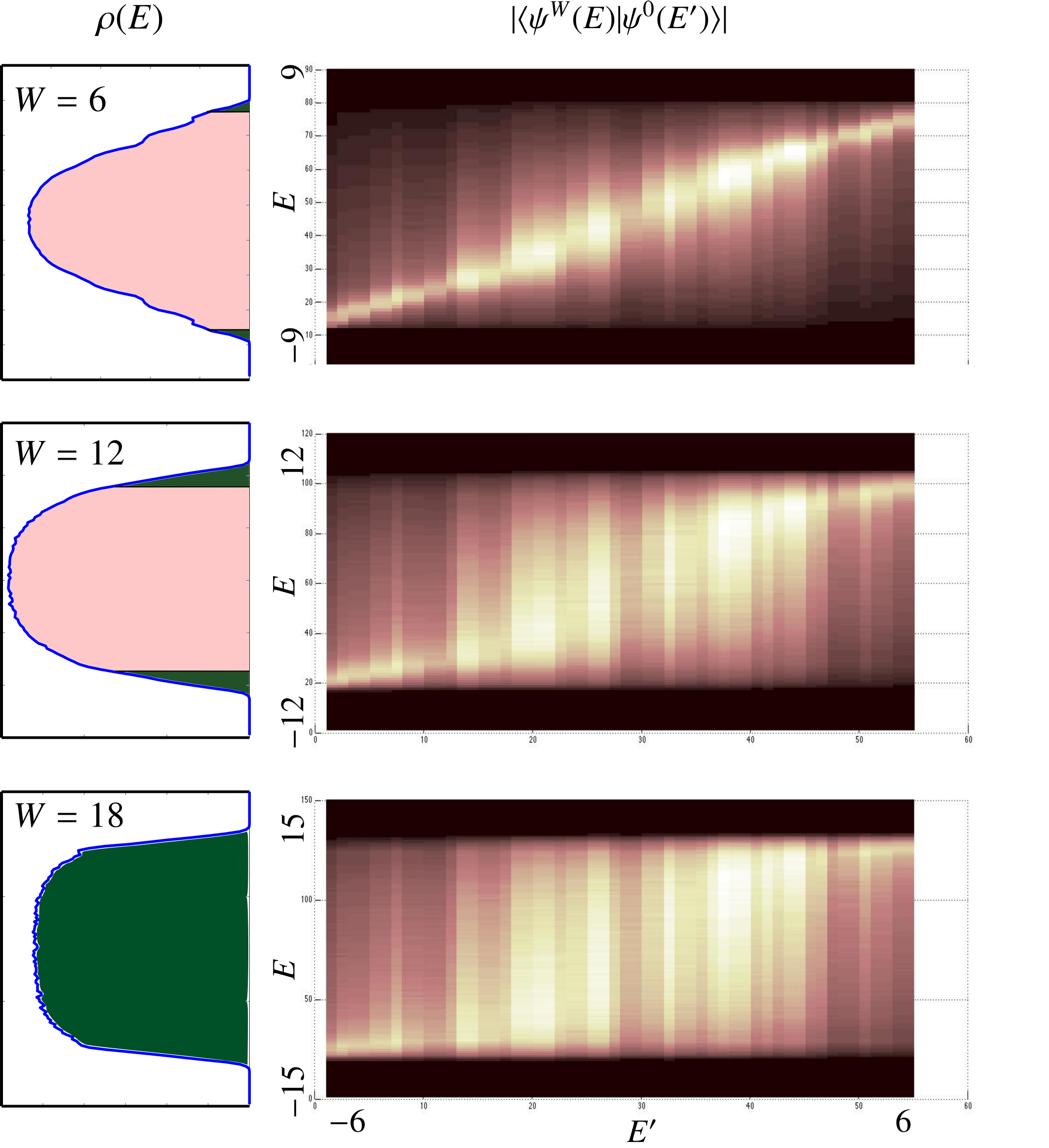}
 \caption{Left: the density of states and the mobility edges separating the dark green and light pink regions for three different strengths of disorder. Right: the magnitudes of the overlaps between these disordered states and plane-wave eigenfunctions of the clean system. Black indicates zero overlap (or no states as the plot contains a factor of the joint density of states) and larger overlaps are shown in lighter color. The expansion of the light region upon increasing $W$ indicates that the overlaps are approaching the uniform distribution of Eq. (11). }
 \label{fig:3}
 \end{center}
\end{figure}

In the left-hand column of Fig.~\ref{fig:3}, we show a few examples of the (disorder-averaged) density of states for the  Hamiltonian of Eq. (\ref{eq:hamil}). The extended (localized) states are shown in light pink (dark green). For any $W>0$ (even for arbitrarily small disorder), localized states appear at the lowest and highest ends of the spectrum for energies $|E|>E_c$ with mobility edges at $\pm E_c$. As we increase $W$, more states become localized. Finally, at $W\approx 16.5$, $E_c\to 0$, the two mobility edges at $\pm E_c$ meet and the full spectrum becomes localized.
In a fermionic system, the many-body ground state is constructed by filling the lowest energy eigenstates up to the Fermi energy $E_F$. When considering the equilibrium ground-state properties, the physics is largely dominated by the nature of the eigenfunction at the Fermi level. If localized, the conductance vanishes and the system is an insulator and if extended, it is a metal. Therefore, the transport properties of the system depends on where the Fermi energy $E_F$ lies in the spectrum with respect to the mobility edges.

In the quench problem studied here, the system is initially in the many-body ground state of the Hamiltonian~(\ref{eq:hamil}) for $W=0$. 
As $N$ many-body wave functions are filled in this fermionic system and each has overlaps with \textit{all} eigenfunctions of the disordered Hamiltonian [shown in Figs.~\ref{fig:2}(b) and \ref{fig:2}(c)], direct detection of the transition in the quench dynamics is challenging. However, the nature and the statistical properties of these wave functions and the distribution of overlaps with the plane waves lead to important crossovers in the behavior of observables after the quench. In particular, as mentioned earlier,  fluctuations of density are suppressed for both extremely localized and extremely extended states but are sensitive to the transient region, where either both extended and localized states are present or there is a large distribution of localization lengths. In what follows, we study this behavior in more depth and show that it results in the nonmonotonic behavior depicted in Fig.~\ref{fig:1}. 

% We note that to avoid ambiguity in the initial state, here we choose the number of particles $N$ such that the many-body ground state is nondegenerate (The many-body wave function is a Slater determinant of single-particle wave functions. For a clean system we have single-particle states of the same energy that form degenerate energy shells. We choose the number of fermions such that no energy shell is partially occupied.) 

\section{WAVE FUNCTION OVERLAPS, OBSERVABLES, AND MOMENTUM DISTRIBUTION}		\label{sec:3}
Here, we present (i) an analysis of wave-function overlaps between pre- and post-quench eigenstates, (ii) a generic formulation for evaluating observables after the quench, and (iii) the behavior of the post-quench momentum distribution.

 The overlaps of the initial and final wave functions plays a central role in quantum quench dynamics; here, an analysis of single-particle wave functions provides most of the required information.  The initial state is a Slater determinant of $N$ plane-wave single-particle wave functions with the lowest energy $\epsilon_{\bf k}$. The final Hamiltonian for each realization of disorder similarly has $p=1\dots L^3$ eigenfunctions $|\psi^p_W\rangle$. At the single-particle level, for any given  initial state $|\psi^n_0\rangle$, the post-quench time-dependent state is governed by the overlap between this state and the eigenstates of the final Hamiltonian.  Specifically, the time-dependence of the wave function can be written as
\begin{equation}
\label{eq:psi0}
\psi^n_0({ {\bf x},t})=\sum_p\langle \psi^p_W|\psi^n_0\rangle e^{-i\varepsilon_W^p t}\psi^p_W({\bf x}),
\end{equation} 
in terms of the overlaps $\langle \psi^p_W|\psi^n_0\rangle$ (throughout the paper we set $\hbar=1$). Clearly, if $W=0$, this overlap is $\delta_{pn}$. For weak disorder, the nature of the wave functions $\psi^p_W({\bf x})$ is determined by the scattering of plane waves off of the disorder potential. To leading order, such scattering mixes plane waves with momenta that are close in energy. Therefore, although the overlaps spread from the delta function above, they remain negligible for states that are far away in energy. As the disorder strength is increased, the disordered wave functions become localized starting at the edges of the spectrum. These localized wave functions are superpositions of a large number of plane waves, which results in an almost uniform distribution of overlaps.

In the right-hand side of Fig. \ref{fig:3}, we show several numerically computed plots of the average overlaps with plane-wave eigenstates of the clean system. The horizontal  axis shows the energy $E'$ of an eigenstate of the clean  Hamiltonian, while the vertical shows $E$, that of the disordered Hamiltonian. Black indicates zero overlap, while lighter colors denotes larger overlaps. As expected, for small disorder, only states close to the diagonal have a large overlap, whereas for large $W$ we approach the situation where each initial plane-wave eigenstate has large overlaps with all the eigenstates of the final Hamiltonian. 

Before discussing the momentum distribution, we formulate the  post-quench dynamics description for generic observables in terms of associated operators and wavefunction overlaps. The initial and final Hamiltonians can be written as $H_0=\Psi^\dagger {\mathscr H}_0 \Psi$ and $H_W=\Psi^\dagger {\mathscr H}_W \Psi$, respectively, where $\Psi^\dagger\equiv(c^\dagger_1\dots c^\dagger_M)$, and ${\mathscr H}_{W}$ is an $M\times M$ Hermitian matrix.
% \begin{equation}
%{\mathscr H}_{W}=\sum_{{\bf x},{\bf y}}{\mathscr H}_{W}^{{\bf xy}}|{\bf x}\rangle\langle {\bf y}|,
%\end{equation}
% where $|{\bf x}\rangle$ represents a quantum state with a single particle on site ${\bf x}$. 
 For one realization of disorder, the quantum expectation value of a quadratic operator
\begin{equation}
O=\Psi^\dagger {\mathscr O}\Psi,
\end{equation}
where ${\mathscr O}$ is an $M\times M$ matrix, can then be computed at time $t$ by writing the Heisenberg operator
\begin{equation}\label{eq:Heis}
O(t)=e^{iH_W t}Oe^{-iH_Wt} =\Psi^\dagger \left(e^{i{\mathscr H}_W t}{\mathscr O}e^{-i{\mathscr H}_Wt} \right)\Psi.
\end{equation}
%The above expression is characteristic of quadratic Hamiltonians and can be derived in many ways, e.g., by using the Baker-Campbell-Hausdorff formula in conjunction with the general relationship $[H_W,O]=\Psi^\dagger [{\mathscr H}_W,{\mathscr O}]\Psi$.
Both clean ($W=0$)and disordered ($W>0$) single-particle Hamiltonians can be diagonalized by a unitary transformation as \begin{equation}\label{eq:diagW}
{\mathscr H}_W=U_WD_WU_W^\dagger,
\end{equation} 
where $D_W={\rm diag}(\varepsilon_W^1\dots \varepsilon_W^M)$, where $\varepsilon  _W^n$ is an eigenvalue of ${\mathscr H}_W$ and the columns of the matrix $U_W$ are the corresponding single-particle eigenfunctions $\psi^n_W({\bf x})$ of the $n$th single-particle level ($\varepsilon_W^n\leqslant \varepsilon_W^{n+1}$).

In terms of quasiparticle operators \begin{equation}
\label{eq:gammaW}
\Gamma^\dagger_W\equiv (\gamma_W^{\dagger 1} \dots \gamma_W^{\dagger M})= \Psi^\dagger U_W,
 \end{equation}
  the Hamiltonians can then be written as $H_W=\Gamma_W^\dagger D_W \Gamma_W=\sum_p \varepsilon_W^p \gamma_W^{\dagger p}\gamma_W^p $. Using Eqs.~(\ref{eq:diagW}) and (\ref{eq:gammaW}), we can then write Eq.~(\ref{eq:Heis}) as
\begin{equation}\label{eq:Heis2}
O(t)=\Gamma_0^\dagger \left(U_0^\dagger U_W e^{iD_Wt}U^\dagger_W{\mathscr O} U_W e^{-iD_W t}U^\dagger_W U_0\right)\Gamma_0,
\end{equation}
where the subscript $0$ indicates $W=0$ in Eq.~(\ref{eq:gammaW}), i.e., $\Gamma^\dagger_0\equiv (\gamma_0^{\dagger 1} \dots \gamma_0^{\dagger M})= \Psi^\dagger U_0$, where the matrix $U_0$ contains the plane-wave eigenfunctions of the clean Hamiltonian [see Eq.~(\ref{eq:psi0})].

As we are working in the Heisenberg picture, we need to take the expectation value of Eq.~(\ref{eq:Heis2})  with the initial many-body state, which is a Fermi sea of $N$ quasi-particles $\Gamma_0$ occupying the lowest energy states:
 \begin{equation}
\label{eq:initial}
|\Psi(0)\rangle=\prod_{n\leqslant N}\gamma^{\dagger n}_0|0\rangle
 \end{equation} 
where $|0\rangle$ is the vacuum. It is easy to observe that only the first $N$ diagonal elements of the $M\times M$ matrix appearing between $\Gamma_0^\dagger$ and $\Gamma_0$ contribute and the quantum expectation value is given by
 \begin{equation}
\langle O(t)\rangle=\sum_{n\leqslant N}\left(U_0^\dagger U_W e^{iD_Wt}U^\dagger_W{\mathscr O} U_W e^{-iD_W t}U^\dagger_W U_0\right)_{nn}.
\end{equation}
The above expression for the quantum expectation value of a general quadratic operator after the quench, can be used as a building block (using the Wick's theorem) for computing the expectation values of higher order operators. In this paper, however, we only discuss quadratic operators.

 Now for the operator $O=c^\dagger_{\bf x}c_{\bf y}$, which will play a role in subsequent discussions, the above equation leads to
\begin{equation}\label{eq:green}
\langle c^\dagger_{\bf x}(t)c_{\bf y}(t)\rangle=\sum_{pq; n\leqslant N}\psi^{*p}_W({\bf x})\psi^q_W({\bf y})\langle\psi^{n}_0|\psi^p_W\rangle\langle\psi^{q}_W|\psi_0^n\rangle
e^{i(\varepsilon_W^p-\varepsilon_W^q)t},
\end{equation}
where the single-particle overlap is given by $\langle\psi^m|\psi^n\rangle\equiv\sum_{\bf x}\psi^m({\bf x})\psi^n({\bf x})$. After some transient time, the expectation value above settles relatively close to its time average, with some persistent temporal fluctuations around it. We refer to this state as a \textit{quasi}-steady state because the temporal fluctuations appear to survive even in the thermodynamic limit. In the next section, focusing on the local density $n_{\bf x}= c^\dagger_{\bf x}c_{\bf x}$, we discuss these fluctuations in more detail.

In evaluating the behavior of any observable  $O$, we have three possible averages to take into account. As discussed above, we have the quantum expectation value (denoted by $\langle O \rangle$) and we assume that this is always taken as the first step. We then have the time average (denoted by an overline), which we take in the long-time limit. For a general time-dependent object $f(t)$, the time average is defined as $\overline{f}\equiv\lim_{T\rightarrow\infty}{1\over T}\int_0^T dt f(t)$. Finally, we have the disorder average taken over many samples and we denote this as $\mathbb{E}(\dots)$, where the dots could be any operator or scalar property of the system (which may or may not depend on time).
% The latter two averages are taken as per the desired measures. 
 We summarize these conventions in the table below:
{\renewcommand{\arraystretch}{1.4}
\begin{center}
\begin{tabular}{ccc}
\hline 
\hline 
Quantum Average & Time Average & Disorder Average\\ 
\hline 
$\langle\dots \rangle$ & $\overline{\dots}$ & $\mathbb{E}(\dots)$\\ 
\hline 
\hline 
\end{tabular} 
\end{center}

We do not expect any spectral degeneracies for a disordered system (with as many random chemical potentials as the number of energy levels). Therefore,
\begin{equation}\label{eq:diag_ave}
\overline{e^{i(\varepsilon_W^p-\varepsilon_W^q)t}}=\delta_{pq}.
\end{equation} 
We now consider the time-averaged behavior of the observable $O=c^\dagger_{\bf x}c_{\bf y}$ of Eq.~(\ref{eq:green}). The only contributions  come from the diagonal terms $p=q$. 
 Hence,  we can write
\begin{equation}\label{eq:ave}
\overline{\langle c^\dagger_{\bf x}c_{\bf y}\rangle}=\sum_{p; n\leqslant N}
\psi^{*p}_W({\bf x})\psi^p_W({\bf y})|\langle\psi^{n}_0|\psi^p_W\rangle|^2,
\end{equation}

Turning to physically motivated situations, the easiest quantity to observe in time-of-flight experiments is the momentum distribution. Here, to present a direct and simple measure for capturing our analysis of wave-function overlaps, we discuss the time and disorder average of the quantum expectation value of the momentum distribution. In a different scenario, where the atomic cloud is released from a trap, signatures of localization in the momentum distribution have been studied in Refs.~\cite{Karpiuk2012,Micklitz2014,Ghosh2014,Lee2014}. Consider the Fourier transform of the fermion annihilation operator 
\begin{equation}
c_{\bf k}={1\over \sqrt{M}}\sum_{\bf x} e^{-i{\bf k}.{\bf x}} c_{\bf x}.
\end{equation}
Using the above expression, the occupation of mode $c_{\bf k}$ is then given by
\begin{equation}\label{eq:mom}
n_{\bf k}\equiv c^\dagger_{\bf k}c_{\bf k}={1\over M}\sum_{{\bf x}{\bf y}}e^{i{\bf k}.{(\bf x}-{\bf y})} c^\dagger_{\bf x}c_{\bf y}.
\end{equation}
The above occupation number of Fourier modes $c_{\bf k}$ can be readily measured in time-of-flight experiments.
%\textbf{{\color{blue}Physically, we need to get the Fourier modes and then compute their occupation number. We can of course mathematically define a direct Fourier transform on $n_{\bf x}$ but it would not have the desired properties like total charge conservation: $\sum_{\bf k}{1\over \sqrt{M}}\sum_{\bf x} e^{-i{\bf k}.{\bf x}} n_{\bf x}=\sqrt{M}n_{x=0}$ while $\sum_{\bf k}{1\over M}\sum_{{\bf x}{\bf y}}e^{i{\bf k}.{(\bf x}-{\bf y})} c^\dagger_{\bf x}c_{\bf y}=\sum_{\bf x}n_{\bf x}$. }}
Using Eqs.~(\ref{eq:mom}) and (\ref{eq:ave}), we have numerically computed the time and quantum averaged $\overline{\langle n_{\bf k}\rangle}$ for each realization of disorder.  We have randomly generated enough realization so that the disorder average $\mathbb{E}(\overline{\langle n_{\bf k}\rangle})$ of this quantity converges in the number of realizations.

The results are shown in Fig.~\ref{fig:4}. As expected for small $W$ the occupation number remains close to the initial Fermi-Dirac distribution. As the overlaps between the eigenstates of the clean and the disordered system become more uniform, the momentum distribution approaches a constant value (equal to the density $N/M$) that is independent of $\bf k$. This can be seen explicitly in the limit of $W\to\infty$, where we can neglect all the hopping terms. When the ratio of hopping to disorder strength approaches $\Gamma/W\to 0$, the probability of $|\Gamma|\ll|\mu_\infty^{\bf x}|$ goes to one  [recall that the hopping amplitude $\Gamma$ was set to unity in Eq.~(\ref{eq:hamil})]. In this strong disorder limit, the wave functions are then localized on individual lattice sites $\psi^p_\infty({\bf x})=\delta_{{\bf x},{\bf x}_p}$. The index $p$ labels the eigenfunctions. As each eigenfunction is localized on one lattice site, there is a one-to-one correspondence between the lattice sites and eigenfunctions so we label the sites with the same index $p$, i.e., $\psi^p_\infty({\bf x})$ is localized on site ${\bf x}_p$. In this limit, we have $\langle\psi_0^n|\psi^p_\infty\rangle={1\over \sqrt{M}}e^{i{\bf k}_n. { \bf x_p}}$, which gives 
\begin{equation}
|\langle\psi_0^n|\psi^p_\infty\rangle|^2=1/M.
\end{equation}
Inserting the above expression into Eq.~(\ref{eq:ave}) then gives 
\begin{equation}\label{eq:cfg}
\overline{\langle c^\dagger_{\bf x}c_{\bf y}\rangle}|_{W\to\infty}={N\over M}\sum_{p}
\psi^{*p}_W({\bf x})\psi^p_W({\bf y})={N\over M}\delta_{\bf xy},
\end{equation}
where we have used the condition of the unitarity. Using Eq.~(\ref{eq:mom}), we then find $\overline{\langle n_{\bf k}\rangle}|_{W\to\infty}=N/M$.

\begin{figure}
\begin{center}
 \includegraphics[width =\columnwidth]{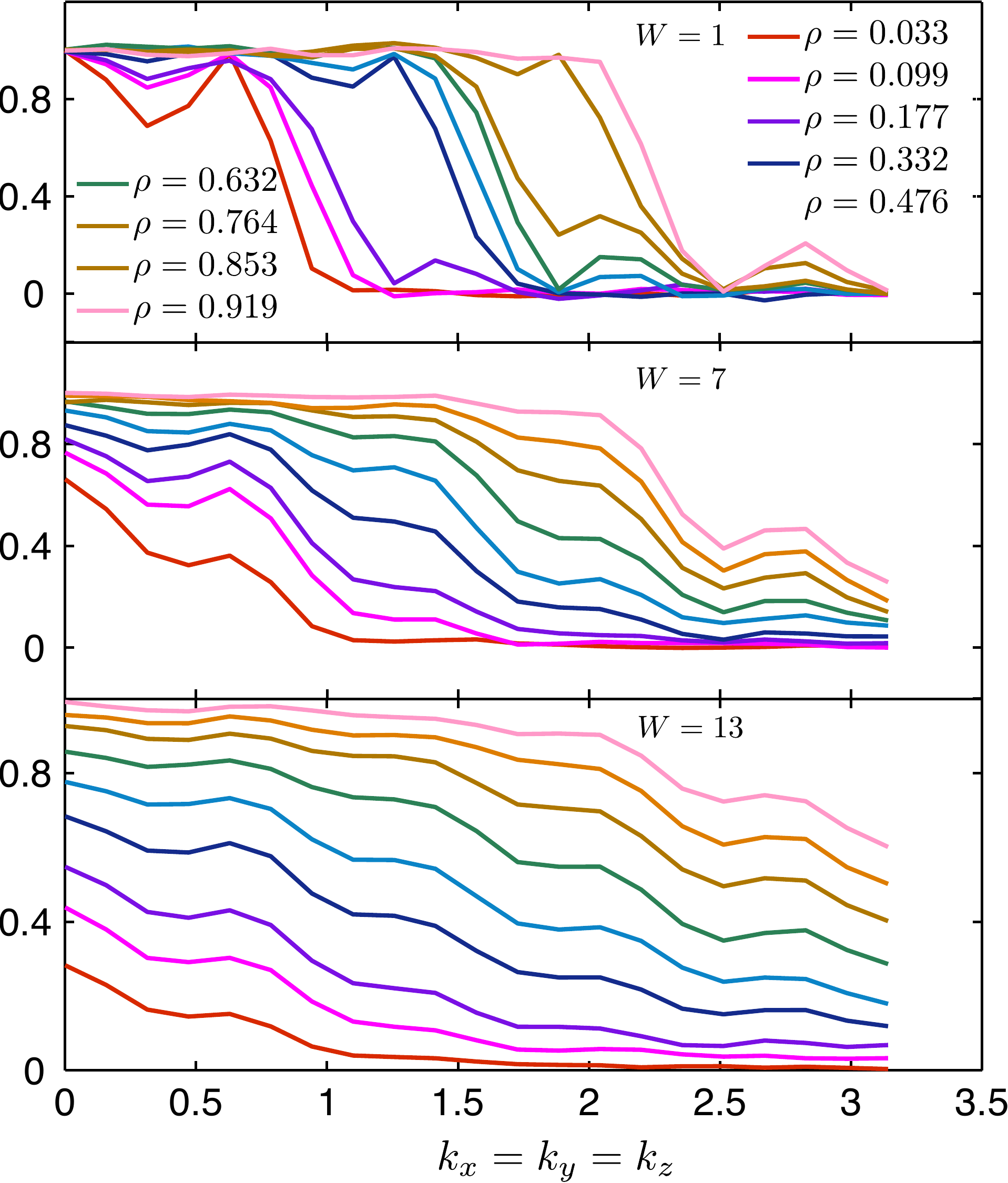}
 \caption{(Color on-line) The long-time-limit disorder-averaged momentum distribution for various densities and strengths of disorder $W$ (in the post-quench Hamiltonian) along a line $k_x=k_y=k_z$ in the  Brillouin zone. Upon increasing $W$, the momentum distribution crosses over from a Fermi-Dirac step function to a uniform distribution.}
 \label{fig:4}
 \end{center}
\end{figure}

While the features of the average momentum distribution reflect localization physics to some degree, we now show that a full-fledged analysis involving temporal and sample-to-sample fluctuations brings out richer effects.

\section{TEMPORAL AND SAMPLE-TO-SAMPLE DENSITY FLUCTUATIONS}	\label{sec:4}
\subsection{Formalism and connection with final eigenstates}\label{sec:4a}
In this section, we consider the fluctuations of local density $n_{\bf x}=c^\dagger_{\bf x} c_{\bf x}$ on a site ${\bf x}$.  Local observables such as $n_{\bf x}$ generically exhibit strong quantum fluctuations characterized by $\langle n_{\bf x}^2\rangle-\langle n_{\bf x}\rangle^2=\langle n_{\bf x}\rangle\left(1-\langle n_{\bf x}\rangle\right)$, where we have made use of the relationship $n_{\bf x}^2=n_{\bf x}$. As the quantum fluctuations are simply related to quantum expectation values $\langle n_{\bf x}\rangle$, here we only focus on \textit{temporal} and \textit{sample-to-sample} fluctuations of these quantum averages as discussed below. The treatment for  temporal fluctuations is similar to those of previous works, which considered fluctuations for the one-dimensional case~\cite{Ziraldo2012,Ziraldo2013}. However, we present results on the dependence of these fluctuations on disorder strength in three dimensions. We also present results on sample-to-sample fluctuations. Our underlying assumption is that for a given sample (i.e., realization of disorder), the quench experiment can be carried out over and over and, thus, the local density at time $t$ after the quantum quench can be measured many times, yielding time- and sample-dependent quantum expectation values $\langle n_{\bf x}(t)\rangle$ (see Fig.~\ref{fig:1}). Moreover, throughout this paper,  we focus on the quasi-steady states reached after the transient de-phasing time scales.

 Generically, local observables $O$ are expected to \textit{equilibrate} at long times, i.e., when $t\rightarrow \infty$, $\langle O(t)\rangle-\overline{\langle O\rangle}\rightarrow 0$.  It has been recently suggested, however, that disordered systems may not equilibrate in the above sense due to their non-smooth spectral properties~\cite{Ziraldo2012,Ziraldo2013}. As the standard notion of thermalization (either to the Gibbs or the generalized Gibbs ensemble) relies on equilibration (the decay of temporal fluctuations), these systems do not thermalize. In the absence of equilibration, we thus have the following hierarchy of fluctuations: (i) quantum fluctuations in a given sample at a fixed time (not discussed further in this paper), (ii) temporal fluctuations of the quantum expectation values around their time average for a typical sample, and (iii) sample-to-sample fluctuations of the above-mentioned time averages. In analogy with the various types of moments used to characterize noise-driven systems~\cite{Dalessio2013}, we characterize the fluctuations (ii) and (iii) respectively by the following moments:
\begin{eqnarray}\label{eq:var1}
{\rm Var}_t[O]&\equiv& \mathbb{E}\left[\overline{\langle O\rangle^2}-\left(\overline{\langle O\rangle}\right)^2\right],\\\label{eq:var2}
{\rm Var}_s[O]&\equiv& \mathbb{E}\left[\left(\overline{\langle O\rangle}\right)^2\right]-\left( \mathbb{E}\left[\overline{\langle O\rangle}\right]\right)^2,
\end{eqnarray}
where various averages are denoted in the table in the previous section. The variation ${\rm Var}_t[O]$ encodes how much the time-dependent $\langle O(t)\rangle$ fluctuates around its time average $\overline{\langle O\rangle}$ for an average sample, while ${\rm Var}_s[O]$ characterizes the fluctuations of the time average $\overline{\langle O\rangle}$ from sample to sample.

To visualize the two types of fluctuations above, we consider the behavior of $\langle n_{\bf x}(t)\rangle$ for different samples as shown  Fig.~\ref{fig:5} (for a system of $L=10$ at half-filling after a sudden quench from $W=0$ to $W=4$ as an example).  The blue circles represent $\langle n_{\bf x}(t)\rangle$ (for a particular site ${\bf x}$) as a function of time. Different data sets correspond to various samples. As seen in the figure, for each sample, $\langle n_{\bf x}(t)\rangle$ keeps fluctuating around its time average $\overline{\langle n_{\bf x}\rangle}$ and does not relax even in the limit of $t\rightarrow \infty$. We mention that we have observed that this behavior persists over time scales that are several orders of magnitude larger than what is shown in the figure. Moreover, the time averages $\overline{\langle n_{\bf x} \rangle}$ (shown in red lines) strongly fluctuate from sample to sample. In the discussion above, we arbitrarily chose a fixed site $\bf x$. With periodic boundary conditions, all sites are equivalent upon disorder averaging and the choice of the site $\bf x$ is unimportant. We note in passing that the sample-to-sample fluctuations are very similar to position-to-position fluctuations in a given sample in the thermodynamic limit.

\begin{figure}
\begin{center}
 \includegraphics[width =\columnwidth]{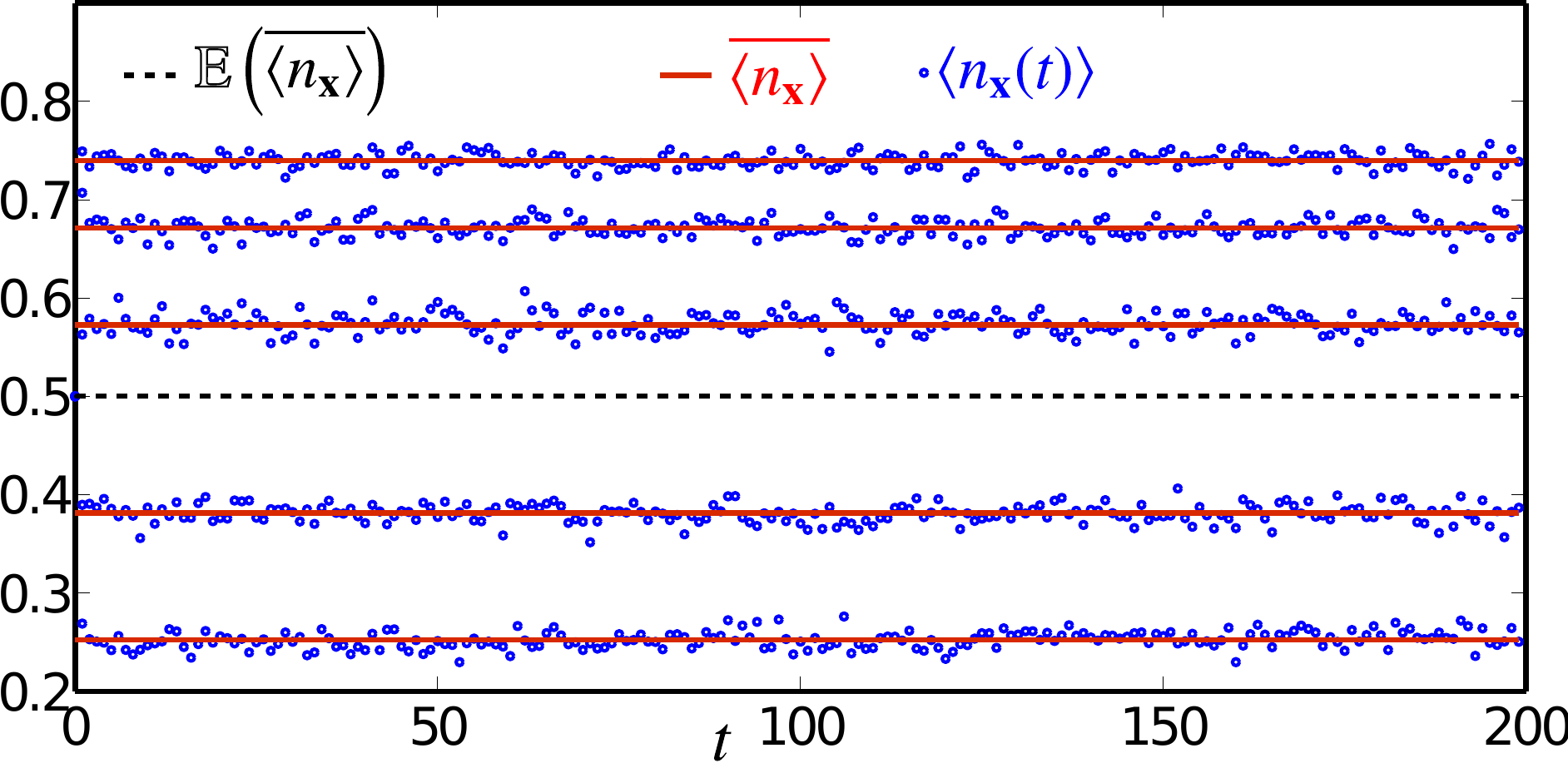}
 \caption{The blue circles represent $\langle n_{\bf x}(t)\rangle$, the density on site ${\bf x}$ at time $t$ after the quench in a system having $M=10^3$, $N=500$, and $W=4$ for various realizations of disorder. The temporal fluctuations around the time averages $\overline{\langle n_{\bf x} \rangle}$ (red solid lines) persist in the limit of $t\rightarrow \infty$ and are characterized by Eq.~(\ref{eq:var1}). The sample-to-sample fluctuations of $\overline{\langle n_{\bf x} \rangle}$ around their average (dashed black line) are characterized by Eq.~(\ref{eq:var2}).}
 \label{fig:5}
 \end{center}
\end{figure}

Before quantifying the fluctuations~(\ref{eq:var1}) and (\ref{eq:var2}), we present a qualitative discussion of the fluctuations. From Eq.~(\ref{eq:green}) for ${\bf x}={\bf y}$, we have 
\begin{equation}\label{eq:ave2}
\langle n_{\bf x}(t)\rangle=\sum_{pq; n\leqslant N}\psi^{*p}_W({\bf x})\psi^q_W({\bf x})\langle\psi^{n}_0|\psi^p_W\rangle\langle\psi^{q}_W|\psi_0^n\rangle
e^{i(\varepsilon_W^p-\varepsilon_W^q)t}.
\end{equation}
From Eq.~(\ref{eq:diag_ave}), it is clear that temporal fluctuations are due to a contribution of states $p$ and 
$q$ with different energies in Eq.~(\ref{eq:ave2}). If $W=0$, there is no quench and everything is stationary. Mathematically, a non-zero product $\langle\psi^{n}_0|\psi^p_W\rangle\langle\psi^{q}_W|\psi_0^n\rangle$ requires $p=q=n$, making the temporal fluctuations vanish. As we increase $W$, we can see from Fig.~\ref{fig:3} that the spreading of the overlaps $\langle\psi^{n}_0|\psi^p_W\rangle$ allows for a larger contribution from states with $\epsilon_p\neq \epsilon_q$. However, the product $\psi^{*p}_W({\bf x})\psi^q_W({\bf x})$ has a competing effect that sets in at very large $W$. As we showed earlier,  $\psi^{*p}_{\infty}({\bf x})=\delta_{{\bf x},{\bf x}_p}$, so when the localization length approaches the lattice spacing, the product $\psi^{*p}_W({\bf x})\psi^q_W({\bf x})$ vanishes for $p\neq q $, which obliterates the temporal fluctuations of Eq.~(\ref{eq:ave2}).

Analogous arguments can be made about the sample-to-sample fluctuations. As mentioned before, we assume that there are no accidental degeneracies in the spectrum of the disordered Hamiltonian and that, therefore, Eqs.~(\ref{eq:diag_ave}) and (\ref{eq:ave}) hold. Now, Eq.~(\ref{eq:ave}) immediately leads to
\begin{equation}\label{eq:mean}
\overline{\langle n_{\bf x}\rangle}=\sum_{p; n\leqslant N}|\psi^{p}_W({\bf x})|^2|\langle\psi^{n}_0|\psi^p_W\rangle|^2.
\end{equation}
Note that after averaging over disorder, the system must exhibit translation invariance and $\mathbb{E}[\overline{\langle n_{\bf x}\rangle}]={1\over M}  \mathbb{E}[\sum_{\bf x}\overline{\langle n_{\bf x}\rangle}]=N/M$ for all ${\bf x}$. We can see this explicitly from Eq.~(\ref{eq:mean}) by using the normalization $\sum_{\bf x}|\psi^{p}_W({\bf x})|^2=1$ and the resolution of identity $\sum_p |\psi_W^p\rangle\langle \psi_W^p|=\mathbb{I}$.

Clearly, for weak disorder, there is little difference between different samples. Increasing $W$ from 0, makes the wave functions for various realizations of disorder different and can increase the variations of expression~(\ref{eq:mean}) from sample-to-sample. However, once again, very large disorder suppresses the fluctuations. For $W\to\infty$ this can be seen from Eq.~(\ref{eq:cfg}), where the time average of the density is $N/M$, independent of the realization of disorder. We argued that the temporal fluctuations vanish when all localization lengths approach the lattice spacing. For sample-to-sample fluctuations, on the other hand, we will later argue that even for relatively large localization length, if there is not much variation in $\xi$, the fluctuations become suppressed. Therefore, after the initial increase as a function of $W$, the sample-to-sample fluctuations begin to decrease at weaker disorder strength in comparison with the temporal fluctuations.

We now proceed to calculate the moments~(\ref{eq:var1}) and (\ref{eq:var2}) (for $O=n_{\bf x}$). To compute the moment~(\ref{eq:var1}), we need $\overline{\langle n_{\bf x}\rangle^2}$, 
 which can be found by using the time average~\cite{Ziraldo2012}
\begin{equation}\label{eq:4ave}
\overline{e^{i(\varepsilon_W^p-\varepsilon_W^q+\varepsilon_W^{p'}-\varepsilon_W^{q'})t}}=\delta_{pq}\delta_{p'q'}+\delta_{pq'}\delta_{p'q}-\delta_{pq}\delta_{qp'}\delta_{p'q'}.
\end{equation}
 In obtaining the expression above, we have assumed that we are working with a finite system of size $L$ in the limit of long times $t\rightarrow \infty$ and then we have taken the limit of large system size. The time average in the equation above is non-vanishing only if the oscillatory term is time-independent, i.e., $\varepsilon_W^p-\varepsilon_W^q+\varepsilon_W^{p'}-\varepsilon_W^{q'}=0$. For a discrete spectrum (corresponding to a finite system), without any accidental degeneracies in energies and \textit{energy gaps}, this can be achieved when $p=q$ and $p'=q'$ or $p=q'$ and $p'=q$, giving rise to the first two terms on the right-hand side of Eq.~(\ref{eq:4ave}). The last term in the equation is added to correct for the over-counting when $p=q=p'=q'$. If the thermodynamic limit $L\rightarrow \infty$ is taken before the limit of $t\rightarrow \infty$  (in the definition of the time average), we do not have a discrete spectrum. For a continuous spectrum, the time average will be non-zero on a three-dimensional plane of the four-dimensional $(pqp'q')$ space characterized by $\varepsilon_W^p-\varepsilon_W^q+\varepsilon_W^{p'}-\varepsilon_W^{q'}=0$, while in the discrete case, the time average is non-zero on a two-dimensional subset of this four-dimensional space. In practice, the order of limits we consider implies that the time scales are much longer than the inverse level spacing of the system.

 We can then write the following expression for one realization of disorder: 
 \begin{equation}\label{eq:temporal}
 \begin{split}
 &\overline{\langle n_{\bf x} \rangle^2}-\left(\overline{\langle n_{\bf x}\rangle}\right)^2=\\
 &\sum_{p\neq q;n,n'\leqslant N}
 |\psi^{p}_W({\bf x})|^2|\psi^q_W({\bf x})|^2\langle \psi^{n}_0|\psi^p_W\rangle\langle\psi^{q}_W|\psi_0^n\rangle
\langle \psi^{n'}_0|\psi^q_W\rangle\langle \psi^{p}_W|\psi_0^{n'}\rangle.
 \end{split} 
\end{equation}
Interestingly, Eqs.~(\ref{eq:mean}) and (\ref{eq:temporal}), which characterize the asymptotic sample-to-sample and temporal density fluctuations through the moments~(\ref{eq:var1}) and (\ref{eq:var2}) (for $O=n_{\bf x}$), are independent of the eigenvalues $\varepsilon_W^p$ and can be be obtained from the statistics of eigenfunctions $\psi_W^p({\bf x})$ alone. Such statistics has been the subject of intensive studies, for e.g., using supersymmetric nonlinear sigma models~\cite{Efetov1997,Mirlin2000}.

Using the translation invariance of the system (upon disorder averaging), we can write both variances ${\rm Var}_{t,s}[n_{\bf x}]$ in a form that is explicitly independent of $\bf x$: ${\rm Var}_{t,s}[n_{\bf x}]={1\over M}\sum_{\bf x} {\rm Var}_{t,s}[n_{\bf x}]$, which leads to
\begin{eqnarray}
{\rm Var}_{t}[n_{\bf x}]=\sum_{\scriptsize \begin{array}{c}
 p\neq q \\ 
n,n'\leqslant N
\end{array}}\mathbb{E}\left[{C^{pq}_W\over M^2}\langle\psi_0^n|\psi^p_W\rangle\langle\psi^{q}_W|\psi_0^n\rangle
\langle \psi^{n'}_0|\psi^q_W\rangle\langle \psi^{p}_W|\psi_0^{n'}\rangle\right],\label{eq:moment1}\\
{\rm Var}_{s}[n_{\bf x}]=\sum_{\scriptsize \begin{array}{c}
 p, q \\ 
n,n'\leqslant N
\end{array}}\mathbb{E}\left[{1\over M^2}\left(C^{pq}_W-1\right)|\langle\psi_0^n|\psi^p_W\rangle|^2
|\langle \psi^{n'}_0|\psi^q_W\rangle|^2\right],~~~~\label{eq:moment2}
\end{eqnarray}
where the two-eigenfunction correlator $C^{pq}_W$ is defined as 
\begin{equation}\label{eq:cpq}
C^{pq}_W\equiv M \sum_{\bf x} |\psi^{p}_W({\bf x})|^2|\psi^{q}_W({\bf x})|^2.
\end{equation}
The statistics of $C^{pq}_W$ has been studied in the context of the statistical properties of disordered eigenfunctions. For $p=q$, it is related to the inverse participation ratio, whose scaling with system size is an important diagnostic for distinguishing localized and extended states. It is easy to observe that for an extended state $p$, $C^{pp}_W$ does not scale with the system size, while for a localized state it scales with $M$. The behavior of $C^{pq}_W$ for two \textit{different} eigenstates has also been studied. If at least one of the eigenstates is extended $C^{pq}_W\approx 1$. If both of the eigenstates are localized, $C^{pq}_W$ vanishes most of the time, except when the localized wave functions overlap. It was shown in Ref.~\cite{Cuevas2007} that on average, we have $\mathbb{E}[C^{pq}_W]\approx 1$ in this case as well.

We can further show that the fluctuations are symmetric under the transformation $N\to M-N$. We first consider ${\rm Var}_{t}[n_{\bf x}]$ with $M-N$ particles:
%\begin{eqnarray}
%{\rm Var}_{t}[n_{\bf x}]|_{N-M}& &=\sum_{\scriptsize \begin{array}{c}
% p\neq q \\ 
%M-N<n,n'
%\end{array}}
%\\
%& &\mathbb{E}\left[{C^{pq}_W\over M^2}
%\langle\psi^{q}_W|\left(\mathbb{I}-|\psi_0^n\rangle\langle\psi_0^n|\right)|\psi^p_W\rangle
%\langle \psi^{p}_W|\left(\mathbb{I}-|\psi_0^{n'}\rangle\langle\psi_0^{n'}|\right)|\psi^q_W\rangle\right],\nonumber
%\end{eqnarray}
\begin{equation}
\begin{split}
{\rm Var}_{t}[n_{\bf x}]|_{N-M}=&\sum_{\scriptsize \begin{array}{c}
 p\neq q \\ 
M-N<n,n'
\end{array}}
\mathbb{E}\bigg[{C^{pq}_W\over M^2}
\langle\psi^{q}_W|\left(\openone-|\psi_0^n\rangle\langle\psi_0^n|\right)\\
&\times|\psi^p_W\rangle
\langle \psi^{p}_W|\left(\openone-|\psi_0^{n'}\rangle\langle\psi_0^{n'}|\right)|\psi^q_W\rangle\bigg],
\end{split}
\end{equation}
where we have used the resolution of identity. Now since $\langle \psi^{p}_W|\mathbb{I}|\psi^q_W\rangle=\delta_{pq}$ and the sum is over $p\neq q$, we find that ${\rm Var}_{t}[n_z]|_{N-M}$ is given by the same expression as ${\rm Var}_{t}[n_z]|_{N}$ except the sums over $n$ and $n'$ are over the $N$ highest-energy wave functions as opposed to the $N$ lowest-energy ones. Noting that the sums over $p$ and $q $ are over all levels and the overlaps $\langle \psi^{n}_0|\psi^p_W\rangle $ on average have a symmetric structure under $E\to -E$ (see Fig.~\ref{fig:3}), we conclude that the fluctuations must be symmetric around half-filling. We can give a similar argument for the sample-to-sample fluctuations again by using the resolution of identity to relate the sum over $M-N$ low-lying states to a sum over the $N$ highest-energy states. Here, we need to show that
\begin{equation}
\sum_{\scriptsize \begin{array}{c}
 p, q \\ 
M-N<n,n'
\end{array}}
\mathbb{E}\left[{1\over M^2}{\left(C^{pq}_W-1\right)}
\langle\psi^{p}_W|\mathbb{I}|\psi^p_W\rangle
\langle \psi^{q}_W|\mathbb{I}|\psi^q_W\rangle\right]=0,
\end{equation}
which follows from $\langle \psi^{p}_W|\mathbb{I}|\psi^p_W\rangle=\langle \psi^{q}_W|\mathbb{I}|\psi^q_W\rangle=1$ and $\sum_{p}\left(C^{pq}_W-1\right)=0$ [see Eq.~(\ref{eq:cpq})].

\subsection{Numerical results and analysis}

Having obtained tractable forms for the temporal and sample-to-sample fluctuations in terms of  the eigenstates of the final Hamiltonian, in this section, we discuss the behavior of these moments using numerical simulations and corroborating analysis.
We first compute the moments~(\ref{eq:moment1}) and (\ref{eq:moment2}) associated with these two fluctuations by direct numerical computation. 
The behavior of these moments as a function of the density $N/M$ is shown in Fig.~\ref{fig:6} for a system size of $L=8$ for quenches terminating in a series of different disorder strengths  $W$. We have obtained good convergence in the disorder-averaged moments by averaging over 1000 samples. For a given disorder strength, we see that both fluctuations increase as a function of density, naturally doing so as more sites become filled. They reach a peak around half filling, and then decrease again as the density increases towards unity, thus allowing fewer and fewer empty sites for fluctuations. The trend holds for all quench disorder strengths. The unique feature that emerges from an interplay between quench dynamics, wave-function overlaps and localization physics, as discussed in previous sections and what follows, is that both fluctuations show non-monotonic behavior as a function of disorder strength.

In Fig.~\ref{fig:7}, we plot the moments at a fixed density (near half-filling for which the maximum occurs) as a function of $W$, which shows a peak at finite $W$. We clearly see the non-monotonic behavior. In addition to the non-monotonic behavior itself, an  important observation is that the peak for temporal fluctuations appears at a much higher $W$. Figure~\ref{fig:7} summarizes the main findings of this work. We provided arguments in Sec.~\ref{sec:4a} for the non-monotonic behavior of both temporal and sample-to-sample fluctuations. To reiterate the salient points, first, by construction, both moments ${\rm Var}_{t,s}[n_{\bf x}]\geqslant0$. We then consider the two extreme cases of extended ($W=0$) and localized ($W\to\infty$) states. As argued in the previous section
\begin{eqnarray}
{\rm Var}_{t,s}[n_{\bf x}]\big|_{W=0}&=&{\rm Var}_{t,s}[n_{\bf x}]\big|_{W\to\infty}=0.
\end{eqnarray}
The above extreme-value calculations immediately imply a nonmonotonic behavior for both moments (unless they identically vanish for all $W$).

To elucidate, considerations of the previous section show that for the sample-to-sample fluctuations, Eq.~(\ref{eq:cfg}) immediately implies that all time-averaged densities are equal to $N/M$ independent of the sample and therefore ${\rm Var}_{s}[n_{\bf x}]\big|_{W\to\infty}=0$ (It is obvious that for $W=0$ all samples are the same and there can not be any sample-to-sample fluctuations). As a check, we find that~(\ref{eq:moment2}) agrees with the above: For $W=0$, we have plane waves~(\ref{eq:pwave}) with $|\psi^{p}_0({\bf x})|^2=1/M$ and, therefore, $C_0^{pq}=1$ so all the terms in sum~(\ref{eq:moment2}) vanish individually. For $W\to\infty$, on the other hand, we have $\psi^p_\infty({\bf x})=\delta_{\bf{x}, {\bf x}_p}$ [state $p$ with wave functin $\psi^p_\infty({\bf x})$ is localized on site ${\bf x}_p$] and it follows from Eq.~(\ref{eq:cpq}) that
\begin{equation}
C_\infty^{pq}\equiv\lim_{W\to\infty}C_W^{pq}=M\delta_{pq}.
\end{equation}
This leads to
\begin{equation}
\lim_{W\to\infty}{\rm Var}_{s}[n_z]={N^2\over M^4}\sum_{pq}\left(M\delta_{pq}-1\right)=0
\end{equation}
Similar considerations can be applied to the temporal fluctuations. In particular, for $W=0$, the overlaps in~(\ref{eq:moment1}) give $\delta_{np} \delta_{qn} \delta_{n'p} \delta_{qn'}$, which vanishes for $p\neq q$ (notice that the sum is over $p\neq q$). For $W\to \infty$, again $C_\infty^{pq}=M\delta_{pq}$, which makes the sum over $p \neq q$ vanish in Eq.~(\ref{eq:moment1}). 
\begin{figure}
\begin{center}
 \includegraphics[width =\columnwidth]{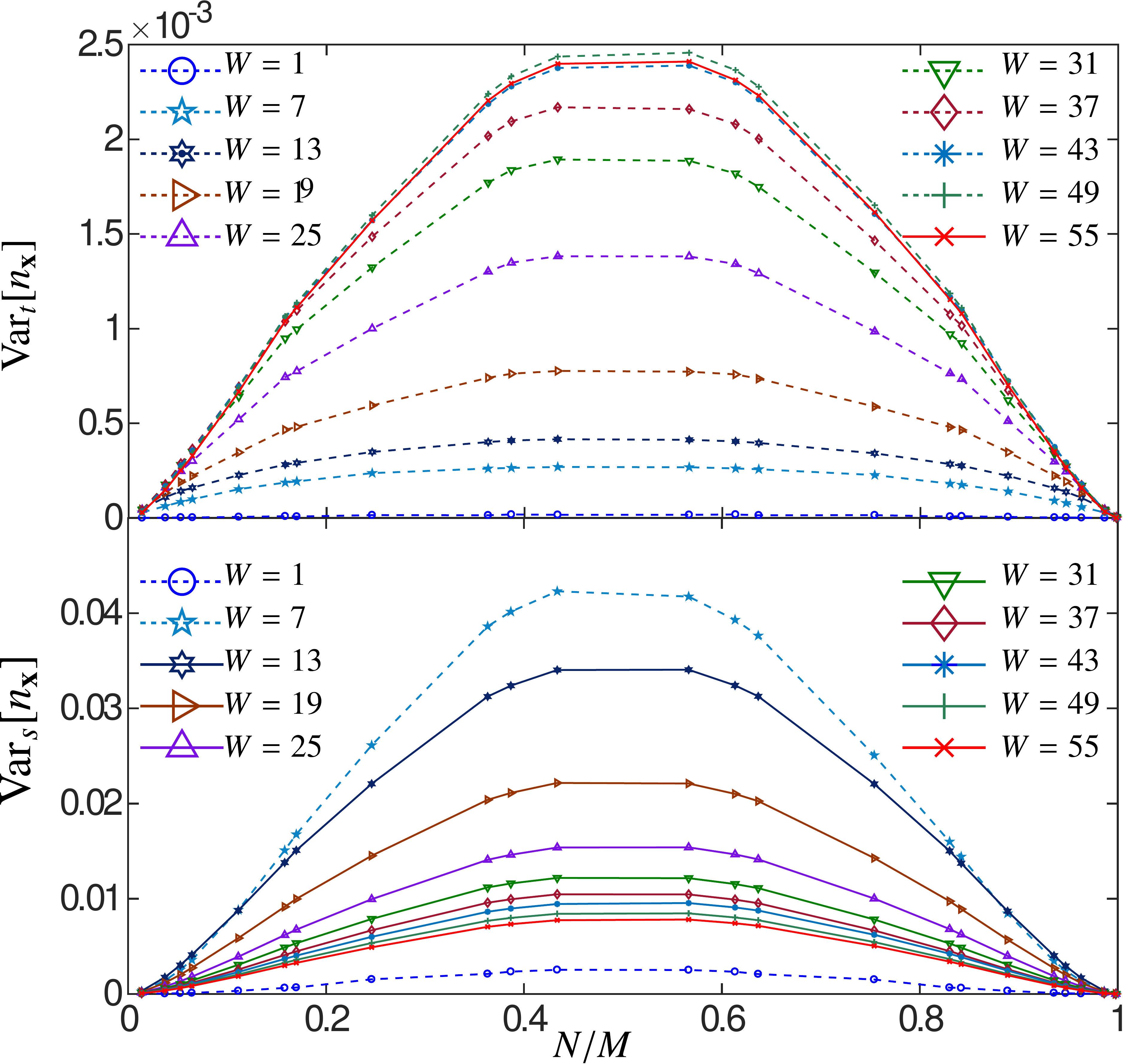}
 \caption{Moments~(\ref{eq:var1}) and~(\ref{eq:var2}) of the local density $n_{\bf x}$ as a function of the average density $N/M$ for various $W$ for $L=8$. The increasing (decreasing) maximum fluctuations as a function of $W$ are indicated by a dashed (solid) line. The apparent imperfect symmetry of the sample-to-sample fluctuations around half filling is an artifact of using a finite number of realizations in averaging over disorder.} 
 \label{fig:6}
 \end{center}
\end{figure}

\begin{figure}
\begin{center}
 \includegraphics[width =.9\columnwidth]{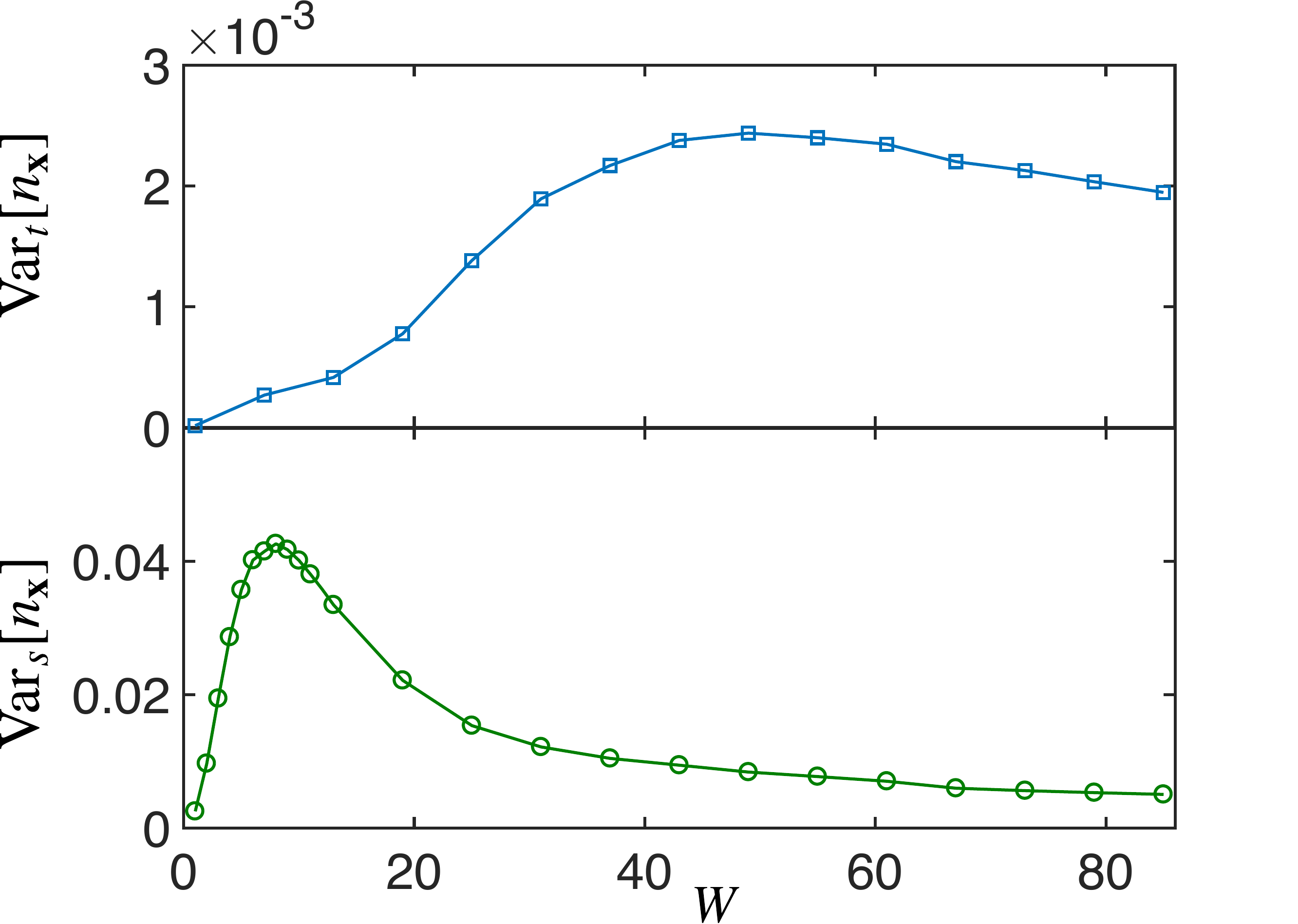}
 \caption{The dependence of the sample-to-sample and temporal fluctuations on $W$ for a fixed density $N/M=0.434$ for $L=8$. Both fluctuations exhibit a nonmonotonic dependence on $W$, first ascending and then descending. However, the temporal fluctuations peak at a much higher $W$ as they rely on reaching localization lengths of the order of the lattice spacing. }
 \label{fig:7}
 \end{center}
\end{figure}

In analyzing the non-monotonicity, the simple argument above does not explain why the peak for the temporal fluctuations appears at a much larger $W$. As mentioned previously, the decreasing sample-to-sample fluctuation relies on the localization lengths becoming uniform (rather than small as in the  case of temporal fluctuations) and therefore sets in at smaller $W$. For large enough disorder, we can assume roughly that the overlaps appearing in the two expressions (\ref{eq:moment1}) and (\ref{eq:moment2}) are uniform and can be factored out of the sum. This is of course a rough approximation but gets better for larger and larger $W$. We can now observe the key difference between the two types of moments. If we assume that the states are localized with a localization $\xi$, $C_W^{pq}$ goes as $M/\xi^3$ with probability $\xi^3/M$ and vanishes otherwise. This indicates that $\mathbb{E}\left(C_W^{pq}-1\right)$, which appears in Eq.~(\ref{eq:moment2}), is not sensitive to the value of $\xi$ and vanishes to leading order. On the other hand, $\mathbb{E}\left(C_W^{pq}\right)$ itself, which appears in Eq.~(\ref{eq:moment1}) does not vanish. The temporal fluctuations become significantly suppressed only when $C_W^{pq}$ approaches the $W\to\infty$ result due to the exclusion of $p=q$ terms in the sum.

Finally, a comment is in order regarding the finite-size effects in the above results. It was shown in Ref.~\cite{Ziraldo2012} that the temporal fluctuations in one dimension eventually saturate to finite values as $L\to\infty$. Here we study this finite-size dependence in the three-dimensional case both for temporal and sample-to-sample fluctuations. In Fig. \ref{fig:8}, we show the two moments for $W=19$ and various system sizes from $L=6$ to $L=14$. The sample-to-sample fluctuations actually increase slightly with system size but quickly saturate the same value. We found that $L=12$ and $L=14$ have very close values of fluctuations for the same density and noise strength. The temporal fluctuations, on the other hand, decrease with increasing system size. 

A similar behavior was observed in Ref.~\cite{Ziraldo2012}, where much larger systems can be studied, but it was found that the temporal fluctuations saturated to finite values. In our three-dimensional system, it is not easy to reach the saturation regime for the temporal fluctuations. As shown in Fig.~\ref{fig:9}, the maximum of ${\rm Var}_t[n_{\bf x}]$ (occurring at half filling) for fixed $W$ fits very well to a quadratic polynomial of $1/L$. The extrapolation based on this quadratic fit is strongly suggestive of the survival of the temporal fluctuations even in the thermodynamic limit.

 The rise and fall  of the two different types of density fluctuations as a function of disorder strength as well as the separation of energy scales for the peak in the temporal and sample-to-sample fluctuations are the key results of this paper. These behaviors are in contrast to quantum quenches in clean systems. They emerge as a subtle  interplay between quench dynamics, wave-function overlaps and localization physics, and capture the manner in which  features of the Anderson localization transition are encoded in the nature of the eigenfunctions and their statistical ensemble. The reduction of temporal fluctuations for strong disorder leads to the observation that increasing disorder could help with observable equilibration.

\begin{figure}
\begin{center}
 \includegraphics[width =\columnwidth]{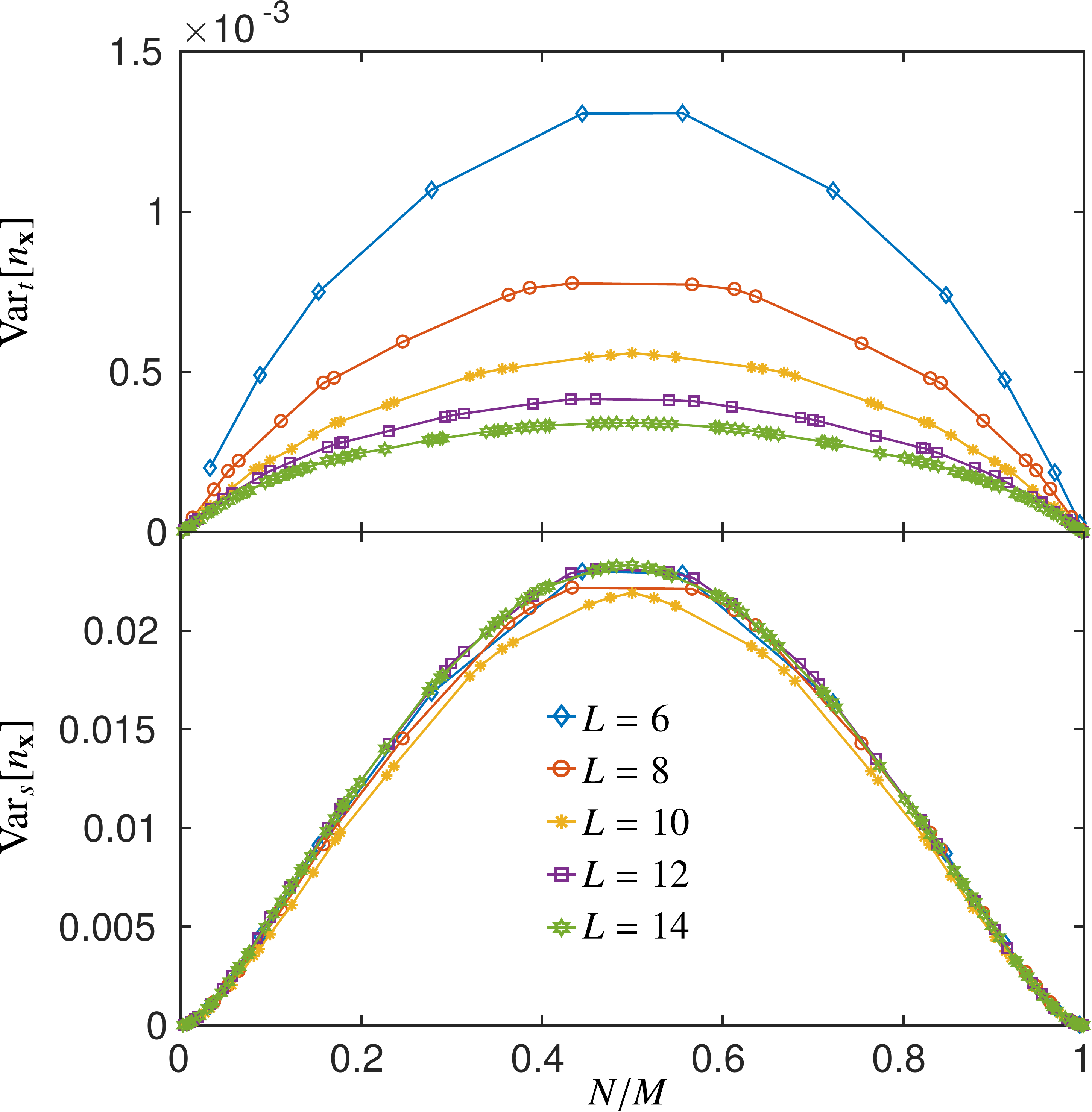}
 \caption{Moments~(\ref{eq:var1}) and~(\ref{eq:var2}) of the local density $n_{\bf x}$ as a function of global average density $N/M$ for various $L$ for $W=19$. The sample-to-sample fluctuations exhibit very weak system-size dependence. The temporal fluctuations slowly decay with system size. Despite small system sizes, extrapolation to $L\to\infty$ suggests survival of these fluctuation in the thermodynamic limit. } 
 \label{fig:8}
 \end{center}
\end{figure}

\begin{figure}
\begin{center}
 \includegraphics[width =\columnwidth]{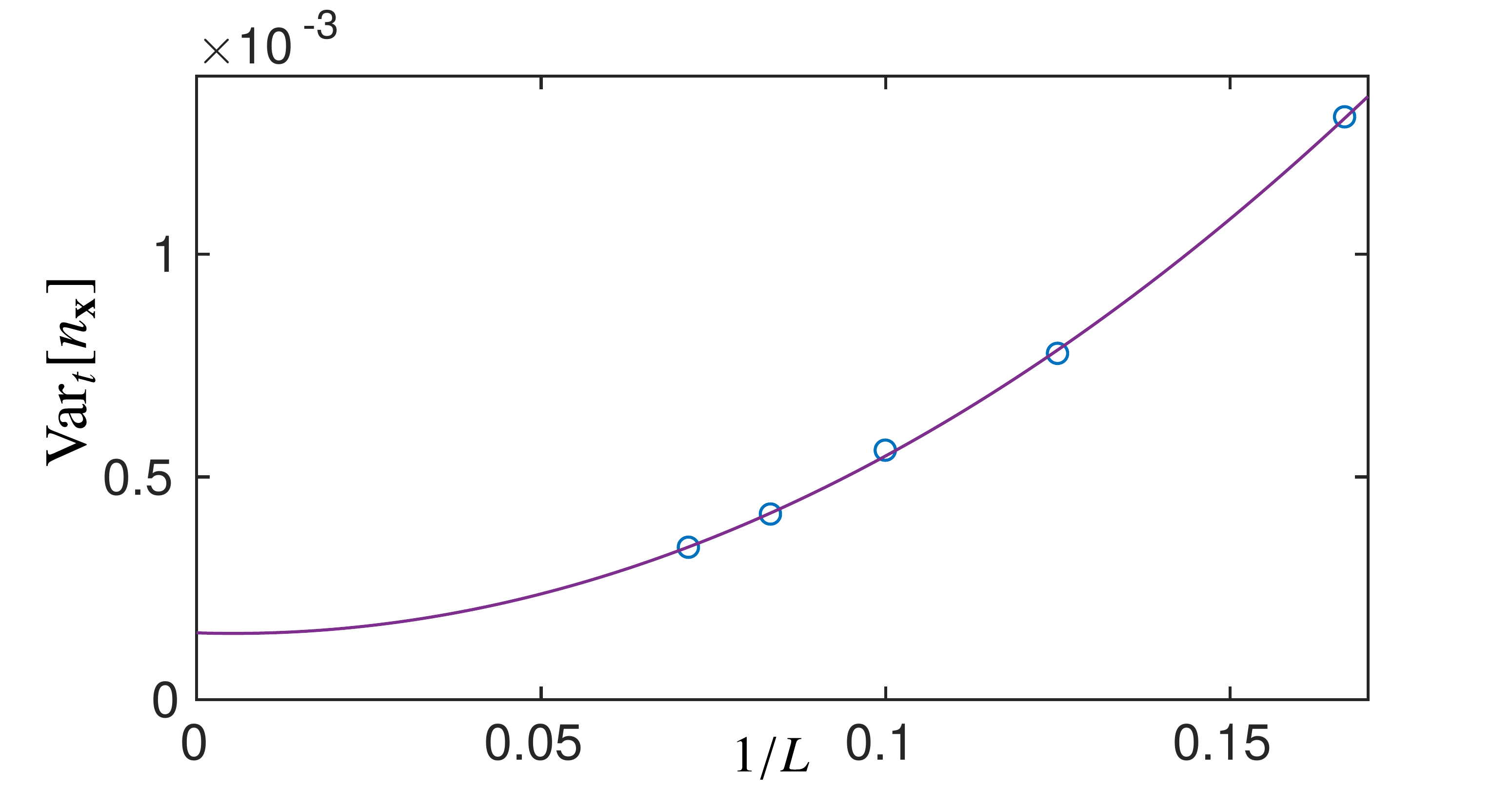}
 \caption{Extrapolation of the maximum value of ${\rm Var}_t[n_{\bf x}]$ (at half-filling) to the thermodynamic limit using a fit to a quadratic function of $1/L$, supporting the survival of these fluctuation for $L\to\infty$. } 
 \label{fig:9}
 \end{center}
\end{figure}
%\begin{figure}
%\begin{center}
% \includegraphics[width =0.8\columnwidth]{scal}
% \caption{(Color on-line) Moments~(\ref{eq:var1}) and~(\ref{eq:var2}) of the local density $n_{\bf x}$ as a function of global average density $N/M$ for various $L$ for $W=19$. The sample-to-sample fluctuations exhibit very weak system-size dependence. The temporal fluctuations slowly decay with system size. Despite small system sizes, extrapolation to $L\to\infty$ suggests survival of these fluctuation in the thermodynamic limit. } 
% \label{fig:9}
% \end{center}
%\end{figure}

\section{Conclusions}\label{sec:5}
In summary,  in a tight-binding model exhibiting Anderson localization, we analyzed a quantum quench consisting of suddenly switching on a disordered potential. While the Anderson transition does not lead to a sharp transition in the resultant post-quench dynamics due to the contribution of many single-particle fermionic levels, the salient features of the transition, namely, the nature and the statistics of the disordered eigenfunctions, give rise to important crossover behaviors. The crossover in the behavior of the overlaps between the final (disordered) and initial (clean) eigenfunctions plays a central role. We demonstrated that as a consequence of the uniform overlaps between localized and extended states, upon increasing the strength of disorder, the momentum distribution at long times after the quench crosses over from the Fermi-Dirac distribution to a constant distribution.

We then turned to the fluctuations of the local density, which constitute the main results of this paper. The persistent temporal fluctuations are a signature of the absence of equilibration in these systems. Interestingly, they survive even in the limit of infinite time and large system sizes. In addition to temporal fluctuations, there are other sources of fluctuations and, in particular, the variations of time averages from sample to sample. Neither of these two types of fluctuations monotonically increases with increasing the strength of disorder. For large disorder strength, they both decrease after some value of disorder strength that depends on the density.

While both fluctuations vanish in the limit of infinite disorder, the temporal fluctuations begin to decrease at a much larger disorder. This is because the reduction in temporal fluctuations relies on reaching regimes where all states are localized with a localization length of the order of the lattice spacing, whereas for the sample-to-sample fluctuations, many extended states and localized states having relatively uniform localization lengths do not contribute to the fluctuations at high enough disorder (where the overlaps between clean and disordered eigenfunctions are roughly uniform).

Our results provide a systematic study of a relatively unexplored problem of the interplay of a system with a Anderson transition and its quench dynamics. They reveal intimate relations between the statistics of the disordered eigenfunctions and post-quench behavior of observables and in particular their fluctuations (which are unique to disordered systems). 

The problem of unitary evolution in disordered systems is especially interesting in light of the recent experimental progress on realizing disordered landscapes and Anderson localization in cold atomic gases~\cite{Sanchez2010,Shapiro2012,Billy2008,Roati2008,Kondov2011,McGehee2013}. The predictions made here would be potentially testable in and highly relevant to such cold atomic settings. For example, the momentum distribution is directly probed by time-of-flight experiments. Our results on the sample-to-sample and temporal fluctuations of the local density can also be observed through \textit{in situ} imaging of the real-space density~\cite{Esteve2006,Gemelke2009,Muller2010,Bakr2010,Sherson2010}. These systems would thus provide an ideal playground for investigating the Anderson-localization physics based on quench dynamics explored in this work.

\section{Acknowledgements}
We are grateful to B. DeMarco, P. Ostrovsky, and M. Rigol for helpful discussions. This work was financially supported by Max Planck-UBC Centre for Quantum Materials (A.R.), Western Washington University (A.R.), and the National Science Foundation under the grant DMR 0644022-CAR (S.V.).  We thank the Kavli Institute for Theoretical Physics for their hospitality and support in initiating this work.

\section{References}

\bibliography{disorder}{}

\begin{thebibliography}{55}%
\makeatletter
\providecommand \@ifxundefined [1]{%
 \@ifx{#1\undefined}
}%
\providecommand \@ifnum [1]{%
 \ifnum #1\expandafter \@firstoftwo
 \else \expandafter \@secondoftwo
 \fi
}%
\providecommand \@ifx [1]{%
 \ifx #1\expandafter \@firstoftwo
 \else \expandafter \@secondoftwo
 \fi
}%
\providecommand \natexlab [1]{#1}%
\providecommand \enquote  [1]{``#1''}%
\providecommand \bibnamefont  [1]{#1}%
\providecommand \bibfnamefont [1]{#1}%
\providecommand \citenamefont [1]{#1}%
\providecommand \href@noop [0]{\@secondoftwo}%
\providecommand \href [0]{\begingroup \@sanitize@url \@href}%
\providecommand \@href[1]{\@@startlink{#1}\@@href}%
\providecommand \@@href[1]{\endgroup#1\@@endlink}%
\providecommand \@sanitize@url [0]{\catcode `\\12\catcode `\$12\catcode
  `\&12\catcode `\#12\catcode `\^12\catcode `\_12\catcode `\%12\relax}%
\providecommand \@@startlink[1]{}%
\providecommand \@@endlink[0]{}%
\providecommand \url  [0]{\begingroup\@sanitize@url \@url }%
\providecommand \@url [1]{\endgroup\@href {#1}{\urlprefix }}%
\providecommand \urlprefix  [0]{URL }%
\providecommand \Eprint [0]{\href }%
\providecommand \doibase [0]{http://dx.doi.org/}%
\providecommand \selectlanguage [0]{\@gobble}%
\providecommand \bibinfo  [0]{\@secondoftwo}%
\providecommand \bibfield  [0]{\@secondoftwo}%
\providecommand \translation [1]{[#1]}%
\providecommand \BibitemOpen [0]{}%
\providecommand \bibitemStop [0]{}%
\providecommand \bibitemNoStop [0]{.\EOS\space}%
\providecommand \EOS [0]{\spacefactor3000\relax}%
\providecommand \BibitemShut  [1]{\csname bibitem#1\endcsname}%
\let\auto@bib@innerbib\@empty
%</preamble>
\bibitem [{\citenamefont {Nandkishore}\ and\ \citenamefont
  {Huse}(2015)}]{Nandkishore2015}%
  \BibitemOpen
  \bibfield  {author} {\bibinfo {author} {\bibfnamefont {R.}~\bibnamefont
  {Nandkishore}}\ and\ \bibinfo {author} {\bibfnamefont {D.~A.}\ \bibnamefont
  {Huse}},\ }\href@noop {} {\bibfield  {journal} {\bibinfo  {journal} {Annu.
  Rev. Condens. Matter Phys.}\ }\textbf {\bibinfo {volume} {6}},\ \bibinfo
  {pages} {15} (\bibinfo {year} {2015})}\BibitemShut {NoStop}%
\bibitem [{\citenamefont {Deutsch}(1991)}]{Deutsch1991}%
  \BibitemOpen
  \bibfield  {author} {\bibinfo {author} {\bibfnamefont {J.~M.}\ \bibnamefont
  {Deutsch}},\ }\href@noop {} {\bibfield  {journal} {\bibinfo  {journal} {Phys.
  Rev. A}\ }\textbf {\bibinfo {volume} {43}},\ \bibinfo {pages} {2046}
  (\bibinfo {year} {1991})}\BibitemShut {NoStop}%
\bibitem [{\citenamefont {Srednicki}(1994)}]{Srednicki1994}%
  \BibitemOpen
  \bibfield  {author} {\bibinfo {author} {\bibfnamefont {M.}~\bibnamefont
  {Srednicki}},\ }\href@noop {} {\bibfield  {journal} {\bibinfo  {journal}
  {Phys. Rev. E}\ }\textbf {\bibinfo {volume} {50}},\ \bibinfo {pages} {888}
  (\bibinfo {year} {1994})}\BibitemShut {NoStop}%
\bibitem [{\citenamefont {Tasaki}(1998)}]{Tasaki1998}%
  \BibitemOpen
  \bibfield  {author} {\bibinfo {author} {\bibfnamefont {H.}~\bibnamefont
  {Tasaki}},\ }\href@noop {} {\bibfield  {journal} {\bibinfo  {journal} {Phys.
  Rev. Lett.}\ }\textbf {\bibinfo {volume} {80}},\ \bibinfo {pages} {1373}
  (\bibinfo {year} {1998})}\BibitemShut {NoStop}%
\bibitem [{\citenamefont {Rigol}\ \emph {et~al.}(2008)\citenamefont {Rigol},
  \citenamefont {Dunjko},\ and\ \citenamefont {Olshanii}}]{Rigol2008}%
  \BibitemOpen
  \bibfield  {author} {\bibinfo {author} {\bibfnamefont {M.}~\bibnamefont
  {Rigol}}, \bibinfo {author} {\bibfnamefont {V.}~\bibnamefont {Dunjko}}, \
  and\ \bibinfo {author} {\bibfnamefont {M.}~\bibnamefont {Olshanii}},\
  }\href@noop {} {\bibfield  {journal} {\bibinfo  {journal} {Nature}\ }\textbf
  {\bibinfo {volume} {452}},\ \bibinfo {pages} {854} (\bibinfo {year}
  {2008})}\BibitemShut {NoStop}%
\bibitem [{\citenamefont {Anderson}(1958)}]{Anderson1958}%
  \BibitemOpen
  \bibfield  {author} {\bibinfo {author} {\bibfnamefont {P.~W.}\ \bibnamefont
  {Anderson}},\ }\href@noop {} {\bibfield  {journal} {\bibinfo  {journal}
  {Phys. Rev.}\ }\textbf {\bibinfo {volume} {109}},\ \bibinfo {pages} {1492}
  (\bibinfo {year} {1958})}\BibitemShut {NoStop}%
\bibitem [{\citenamefont {Basko}\ \emph {et~al.}(2006)\citenamefont {Basko},
  \citenamefont {Aleiner},\ and\ \citenamefont {Altshuler}}]{Basko2006}%
  \BibitemOpen
  \bibfield  {author} {\bibinfo {author} {\bibfnamefont {D.~M.}\ \bibnamefont
  {Basko}}, \bibinfo {author} {\bibfnamefont {I.~L.}\ \bibnamefont {Aleiner}},
  \ and\ \bibinfo {author} {\bibfnamefont {B.~L.}\ \bibnamefont {Altshuler}},\
  }\href@noop {} {\bibfield  {journal} {\bibinfo  {journal} {Ann. Phys.}\
  }\textbf {\bibinfo {volume} {321}},\ \bibinfo {pages} {1126} (\bibinfo {year}
  {2006})}\BibitemShut {NoStop}%
\bibitem [{\citenamefont {Oganesyan}\ and\ \citenamefont
  {Huse}(2007)}]{Oganesyan2007}%
  \BibitemOpen
  \bibfield  {author} {\bibinfo {author} {\bibfnamefont {V.}~\bibnamefont
  {Oganesyan}}\ and\ \bibinfo {author} {\bibfnamefont {D.~A.}\ \bibnamefont
  {Huse}},\ }\href@noop {} {\bibfield  {journal} {\bibinfo  {journal} {Phys.
  Rev. B}\ }\textbf {\bibinfo {volume} {75}},\ \bibinfo {pages} {155111}
  (\bibinfo {year} {2007})}\BibitemShut {NoStop}%
\bibitem [{\citenamefont {Pal}\ and\ \citenamefont {Huse}(2010)}]{Pal2010}%
  \BibitemOpen
  \bibfield  {author} {\bibinfo {author} {\bibfnamefont {A.}~\bibnamefont
  {Pal}}\ and\ \bibinfo {author} {\bibfnamefont {D.~A.}\ \bibnamefont {Huse}},\
  }\href@noop {} {\bibfield  {journal} {\bibinfo  {journal} {Phys. Rev. B}\
  }\textbf {\bibinfo {volume} {82}},\ \bibinfo {pages} {174411} (\bibinfo
  {year} {2010})}\BibitemShut {NoStop}%
\bibitem [{\citenamefont {Polkovnikov}\ \emph {et~al.}(2011)\citenamefont
  {Polkovnikov}, \citenamefont {Sengupta}, \citenamefont {Silva},\ and\
  \citenamefont {Vengalattore}}]{Polkovnikov2011}%
  \BibitemOpen
  \bibfield  {author} {\bibinfo {author} {\bibfnamefont {A.}~\bibnamefont
  {Polkovnikov}}, \bibinfo {author} {\bibfnamefont {K.}~\bibnamefont
  {Sengupta}}, \bibinfo {author} {\bibfnamefont {A.}~\bibnamefont {Silva}}, \
  and\ \bibinfo {author} {\bibfnamefont {M.}~\bibnamefont {Vengalattore}},\
  }\href@noop {} {\bibfield  {journal} {\bibinfo  {journal} {Rev. Mod. Phys.}\
  }\textbf {\bibinfo {volume} {83}},\ \bibinfo {pages} {863} (\bibinfo {year}
  {2011})}\BibitemShut {NoStop}%
\bibitem [{\citenamefont {Greiner}\ \emph {et~al.}(2002)\citenamefont
  {Greiner}, \citenamefont {Mandel}, \citenamefont {H\"ansch},\ and\
  \citenamefont {Bloch}}]{Greiner2002}%
  \BibitemOpen
  \bibfield  {author} {\bibinfo {author} {\bibfnamefont {M.}~\bibnamefont
  {Greiner}}, \bibinfo {author} {\bibfnamefont {O.}~\bibnamefont {Mandel}},
  \bibinfo {author} {\bibfnamefont {T.~W.}\ \bibnamefont {H\"ansch}}, \ and\
  \bibinfo {author} {\bibfnamefont {I.}~\bibnamefont {Bloch}},\ }\href@noop {}
  {\bibfield  {journal} {\bibinfo  {journal} {Nature}\ }\textbf {\bibinfo
  {volume} {419}},\ \bibinfo {pages} {51} (\bibinfo {year} {2002})}\BibitemShut
  {NoStop}%
\bibitem [{\citenamefont {Kinoshita}\ \emph {et~al.}(2006)\citenamefont
  {Kinoshita}, \citenamefont {Wenger},\ and\ \citenamefont
  {Weiss}}]{Kinoshita2006}%
  \BibitemOpen
  \bibfield  {author} {\bibinfo {author} {\bibfnamefont {T.}~\bibnamefont
  {Kinoshita}}, \bibinfo {author} {\bibfnamefont {T.}~\bibnamefont {Wenger}}, \
  and\ \bibinfo {author} {\bibfnamefont {D.~S.}\ \bibnamefont {Weiss}},\
  }\href@noop {} {\bibfield  {journal} {\bibinfo  {journal} {Nature}\ }\textbf
  {\bibinfo {volume} {440}},\ \bibinfo {pages} {900} (\bibinfo {year}
  {2006})}\BibitemShut {NoStop}%
\bibitem [{\citenamefont {Calabrese}\ and\ \citenamefont
  {Cardy}(2006)}]{Calabrese2006}%
  \BibitemOpen
  \bibfield  {author} {\bibinfo {author} {\bibfnamefont {P.}~\bibnamefont
  {Calabrese}}\ and\ \bibinfo {author} {\bibfnamefont {J.}~\bibnamefont
  {Cardy}},\ }\href@noop {} {\bibfield  {journal} {\bibinfo  {journal} {Phys.
  Rev. Lett.}\ }\textbf {\bibinfo {volume} {96}},\ \bibinfo {pages} {136801}
  (\bibinfo {year} {2006})}\BibitemShut {NoStop}%
\bibitem [{\citenamefont {Kollath}\ \emph {et~al.}(2007)\citenamefont
  {Kollath}, \citenamefont {L\"auchli},\ and\ \citenamefont
  {Altman}}]{Kollath2007}%
  \BibitemOpen
  \bibfield  {author} {\bibinfo {author} {\bibfnamefont {C.}~\bibnamefont
  {Kollath}}, \bibinfo {author} {\bibfnamefont {A.~M.}\ \bibnamefont
  {L\"auchli}}, \ and\ \bibinfo {author} {\bibfnamefont {E.}~\bibnamefont
  {Altman}},\ }\href@noop {} {\bibfield  {journal} {\bibinfo  {journal} {Phys.
  Rev. Lett.}\ }\textbf {\bibinfo {volume} {98}},\ \bibinfo {pages} {180601}
  (\bibinfo {year} {2007})}\BibitemShut {NoStop}%
\bibitem [{\citenamefont {Cramer}\ \emph {et~al.}(2008)\citenamefont {Cramer},
  \citenamefont {Dawson}, \citenamefont {Eisert},\ and\ \citenamefont
  {Osborne}}]{Cramer2008}%
  \BibitemOpen
  \bibfield  {author} {\bibinfo {author} {\bibfnamefont {M.}~\bibnamefont
  {Cramer}}, \bibinfo {author} {\bibfnamefont {C.~M.}\ \bibnamefont {Dawson}},
  \bibinfo {author} {\bibfnamefont {J.}~\bibnamefont {Eisert}}, \ and\ \bibinfo
  {author} {\bibfnamefont {T.~J.}\ \bibnamefont {Osborne}},\ }\href@noop {}
  {\bibfield  {journal} {\bibinfo  {journal} {Phys. Rev. Lett.}\ }\textbf
  {\bibinfo {volume} {100}},\ \bibinfo {pages} {030602} (\bibinfo {year}
  {2008})}\BibitemShut {NoStop}%
\bibitem [{\citenamefont {Yukalov}(2011)}]{Yukalov2011}%
  \BibitemOpen
  \bibfield  {author} {\bibinfo {author} {\bibfnamefont {V.}~\bibnamefont
  {Yukalov}},\ }\href@noop {} {\bibfield  {journal} {\bibinfo  {journal} {Laser
  Phys. Lett.}\ }\textbf {\bibinfo {volume} {8}},\ \bibinfo {pages} {485}
  (\bibinfo {year} {2011})}\BibitemShut {NoStop}%
\bibitem [{\citenamefont {Eisert}(2014)}]{Eisert2014}%
  \BibitemOpen
  \bibfield  {author} {\bibinfo {author} {\bibfnamefont {J.}~\bibnamefont
  {Eisert}},\ }\href@noop {} {\bibfield  {journal} {\bibinfo  {journal} {Nat.
  Phys}\ }\textbf {\bibinfo {volume} {11}},\ \bibinfo {pages} {124} (\bibinfo
  {year} {2014})}\BibitemShut {NoStop}%
\bibitem [{\citenamefont {Canovi}\ \emph {et~al.}(2011)\citenamefont {Canovi},
  \citenamefont {Rossini}, \citenamefont {Fazio}, \citenamefont {Santoro},\
  and\ \citenamefont {Silva}}]{Canovi2011}%
  \BibitemOpen
  \bibfield  {author} {\bibinfo {author} {\bibfnamefont {E.}~\bibnamefont
  {Canovi}}, \bibinfo {author} {\bibfnamefont {D.}~\bibnamefont {Rossini}},
  \bibinfo {author} {\bibfnamefont {R.}~\bibnamefont {Fazio}}, \bibinfo
  {author} {\bibfnamefont {G.~E.}\ \bibnamefont {Santoro}}, \ and\ \bibinfo
  {author} {\bibfnamefont {A.}~\bibnamefont {Silva}},\ }\href@noop {}
  {\bibfield  {journal} {\bibinfo  {journal} {Phys. Rev. B}\ }\textbf {\bibinfo
  {volume} {83}},\ \bibinfo {pages} {094431} (\bibinfo {year}
  {2011})}\BibitemShut {NoStop}%
\bibitem [{\citenamefont {Bardarson}\ \emph {et~al.}(2012)\citenamefont
  {Bardarson}, \citenamefont {Pollmann},\ and\ \citenamefont
  {Moore}}]{Bardarson2012}%
  \BibitemOpen
  \bibfield  {author} {\bibinfo {author} {\bibfnamefont {J.~H.}\ \bibnamefont
  {Bardarson}}, \bibinfo {author} {\bibfnamefont {F.}~\bibnamefont {Pollmann}},
  \ and\ \bibinfo {author} {\bibfnamefont {J.~E.}\ \bibnamefont {Moore}},\
  }\href@noop {} {\bibfield  {journal} {\bibinfo  {journal} {Phys. Rev. Lett.}\
  }\textbf {\bibinfo {volume} {109}},\ \bibinfo {pages} {017202} (\bibinfo
  {year} {2012})}\BibitemShut {NoStop}%
\bibitem [{\citenamefont {Chandran}\ \emph {et~al.}(2013)\citenamefont
  {Chandran}, \citenamefont {Nanduri}, \citenamefont {Gubser},\ and\
  \citenamefont {Sondhi}}]{Chandran2013}%
  \BibitemOpen
  \bibfield  {author} {\bibinfo {author} {\bibfnamefont {A.}~\bibnamefont
  {Chandran}}, \bibinfo {author} {\bibfnamefont {A.}~\bibnamefont {Nanduri}},
  \bibinfo {author} {\bibfnamefont {S.~S.}\ \bibnamefont {Gubser}}, \ and\
  \bibinfo {author} {\bibfnamefont {S.~L.}\ \bibnamefont {Sondhi}},\
  }\href@noop {} {\bibfield  {journal} {\bibinfo  {journal} {Phys. Rev. B}\
  }\textbf {\bibinfo {volume} {88}},\ \bibinfo {pages} {024306} (\bibinfo
  {year} {2013})}\BibitemShut {NoStop}%
\bibitem [{\citenamefont {Zangara}\ \emph {et~al.}(2013)\citenamefont
  {Zangara}, \citenamefont {Dente}, \citenamefont {Iucci}, \citenamefont
  {Levstein},\ and\ \citenamefont {Pastawski}}]{Zangara2013}%
  \BibitemOpen
  \bibfield  {author} {\bibinfo {author} {\bibfnamefont {P.~R.}\ \bibnamefont
  {Zangara}}, \bibinfo {author} {\bibfnamefont {A.~D.}\ \bibnamefont {Dente}},
  \bibinfo {author} {\bibfnamefont {A.}~\bibnamefont {Iucci}}, \bibinfo
  {author} {\bibfnamefont {P.~R.}\ \bibnamefont {Levstein}}, \ and\ \bibinfo
  {author} {\bibfnamefont {H.~M.}\ \bibnamefont {Pastawski}},\ }\href@noop {}
  {\bibfield  {journal} {\bibinfo  {journal} {Phys. Rev. B}\ }\textbf {\bibinfo
  {volume} {88}},\ \bibinfo {pages} {195106} (\bibinfo {year}
  {2013})}\BibitemShut {NoStop}%
\bibitem [{\citenamefont {Vosk}\ and\ \citenamefont {Altman}(2013)}]{Vosk2013}%
  \BibitemOpen
  \bibfield  {author} {\bibinfo {author} {\bibfnamefont {R.}~\bibnamefont
  {Vosk}}\ and\ \bibinfo {author} {\bibfnamefont {E.}~\bibnamefont {Altman}},\
  }\href@noop {} {\bibfield  {journal} {\bibinfo  {journal} {Phys. Rev. Lett.}\
  }\textbf {\bibinfo {volume} {110}},\ \bibinfo {pages} {067204} (\bibinfo
  {year} {2013})}\BibitemShut {NoStop}%
\bibitem [{\citenamefont {Sorg}\ \emph {et~al.}(2014)\citenamefont {Sorg},
  \citenamefont {Vidmar}, \citenamefont {Pollet},\ and\ \citenamefont
  {Heidrich-Meisner}}]{Sorg2014}%
  \BibitemOpen
  \bibfield  {author} {\bibinfo {author} {\bibfnamefont {S.}~\bibnamefont
  {Sorg}}, \bibinfo {author} {\bibfnamefont {L.}~\bibnamefont {Vidmar}},
  \bibinfo {author} {\bibfnamefont {L.}~\bibnamefont {Pollet}}, \ and\ \bibinfo
  {author} {\bibfnamefont {F.}~\bibnamefont {Heidrich-Meisner}},\ }\href@noop
  {} {\bibfield  {journal} {\bibinfo  {journal} {Phys. Rev. A}\ }\textbf
  {\bibinfo {volume} {90}},\ \bibinfo {pages} {033606} (\bibinfo {year}
  {2014})}\BibitemShut {NoStop}%
\bibitem [{\citenamefont {Tang}\ \emph {et~al.}(2015)\citenamefont {Tang},
  \citenamefont {Iyer},\ and\ \citenamefont {Rigol}}]{Tang2015}%
  \BibitemOpen
  \bibfield  {author} {\bibinfo {author} {\bibfnamefont {B.}~\bibnamefont
  {Tang}}, \bibinfo {author} {\bibfnamefont {D.}~\bibnamefont {Iyer}}, \ and\
  \bibinfo {author} {\bibfnamefont {M.}~\bibnamefont {Rigol}},\ }\href@noop {}
  {\bibfield  {journal} {\bibinfo  {journal} {Phys. Rev. B}\ }\textbf {\bibinfo
  {volume} {91}},\ \bibinfo {pages} {161109} (\bibinfo {year}
  {2015})}\BibitemShut {NoStop}%
\bibitem [{\citenamefont {Ziraldo}\ \emph {et~al.}(2012)\citenamefont
  {Ziraldo}, \citenamefont {Silva},\ and\ \citenamefont
  {Santoro}}]{Ziraldo2012}%
  \BibitemOpen
  \bibfield  {author} {\bibinfo {author} {\bibfnamefont {S.}~\bibnamefont
  {Ziraldo}}, \bibinfo {author} {\bibfnamefont {A.}~\bibnamefont {Silva}}, \
  and\ \bibinfo {author} {\bibfnamefont {G.~E.}\ \bibnamefont {Santoro}},\
  }\href@noop {} {\bibfield  {journal} {\bibinfo  {journal} {Phys. Rev. Lett.}\
  }\textbf {\bibinfo {volume} {109}},\ \bibinfo {pages} {247205} (\bibinfo
  {year} {2012})}\BibitemShut {NoStop}%
\bibitem [{\citenamefont {Ziraldo}\ and\ \citenamefont
  {Santoro}(2013)}]{Ziraldo2013}%
  \BibitemOpen
  \bibfield  {author} {\bibinfo {author} {\bibfnamefont {S.}~\bibnamefont
  {Ziraldo}}\ and\ \bibinfo {author} {\bibfnamefont {G.~E.}\ \bibnamefont
  {Santoro}},\ }\href@noop {} {\bibfield  {journal} {\bibinfo  {journal} {Phys.
  Rev. B}\ }\textbf {\bibinfo {volume} {87}},\ \bibinfo {pages} {064201}
  (\bibinfo {year} {2013})}\BibitemShut {NoStop}%
\bibitem [{\citenamefont {Gramsch}\ and\ \citenamefont
  {Rigol}(2012)}]{Gramsch2012}%
  \BibitemOpen
  \bibfield  {author} {\bibinfo {author} {\bibfnamefont {C.}~\bibnamefont
  {Gramsch}}\ and\ \bibinfo {author} {\bibfnamefont {M.}~\bibnamefont
  {Rigol}},\ }\href@noop {} {\bibfield  {journal} {\bibinfo  {journal} {Phys.
  Rev. A}\ }\textbf {\bibinfo {volume} {86}},\ \bibinfo {pages} {053615}
  (\bibinfo {year} {2012})}\BibitemShut {NoStop}%
\bibitem [{\citenamefont {Ribeiro}\ \emph {et~al.}(2013)\citenamefont
  {Ribeiro}, \citenamefont {Haque},\ and\ \citenamefont
  {Lazarides}}]{Ribeiro2013}%
  \BibitemOpen
  \bibfield  {author} {\bibinfo {author} {\bibfnamefont {P.}~\bibnamefont
  {Ribeiro}}, \bibinfo {author} {\bibfnamefont {M.}~\bibnamefont {Haque}}, \
  and\ \bibinfo {author} {\bibfnamefont {A.}~\bibnamefont {Lazarides}},\
  }\href@noop {} {\bibfield  {journal} {\bibinfo  {journal} {Phys. Rev. A}\
  }\textbf {\bibinfo {volume} {87}},\ \bibinfo {pages} {043635} (\bibinfo
  {year} {2013})}\BibitemShut {NoStop}%
\bibitem [{\citenamefont {He}\ \emph {et~al.}(2013)\citenamefont {He},
  \citenamefont {Santos}, \citenamefont {Wright},\ and\ \citenamefont
  {Rigol}}]{He2013}%
  \BibitemOpen
  \bibfield  {author} {\bibinfo {author} {\bibfnamefont {K.}~\bibnamefont
  {He}}, \bibinfo {author} {\bibfnamefont {L.~F.}\ \bibnamefont {Santos}},
  \bibinfo {author} {\bibfnamefont {T.~M.}\ \bibnamefont {Wright}}, \ and\
  \bibinfo {author} {\bibfnamefont {M.}~\bibnamefont {Rigol}},\ }\href@noop {}
  {\bibfield  {journal} {\bibinfo  {journal} {Phys. Rev. A}\ }\textbf {\bibinfo
  {volume} {87}},\ \bibinfo {pages} {063637} (\bibinfo {year}
  {2013})}\BibitemShut {NoStop}%
\bibitem [{\citenamefont {Wang}\ and\ \citenamefont {Tong}()}]{Wang2017}%
  \BibitemOpen
  \bibfield  {author} {\bibinfo {author} {\bibfnamefont {X.}~\bibnamefont
  {Wang}}\ and\ \bibinfo {author} {\bibfnamefont {P.}~\bibnamefont {Tong}},\
  }\href@noop {} {\bibinfo  {journal} {J. Stat Phys.: Theor. Exp. (2017)
  113107}\ }\BibitemShut {NoStop}%
\bibitem [{\citenamefont {Mirlin}(2000)}]{Mirlin2000}%
  \BibitemOpen
\bibfield  {journal} {  }\bibfield  {author} {\bibinfo {author} {\bibfnamefont
  {A.~D.}\ \bibnamefont {Mirlin}},\ }\href@noop {} {\bibfield  {journal}
  {\bibinfo  {journal} {Phys. Rep.}\ }\textbf {\bibinfo {volume} {326}},\
  \bibinfo {pages} {259 } (\bibinfo {year} {2000})}\BibitemShut {NoStop}%
\bibitem [{\citenamefont {Bloch}\ \emph {et~al.}(2008)\citenamefont {Bloch},
  \citenamefont {Dalibard},\ and\ \citenamefont {Zwerger}}]{Bloch2008}%
  \BibitemOpen
  \bibfield  {author} {\bibinfo {author} {\bibfnamefont {I.}~\bibnamefont
  {Bloch}}, \bibinfo {author} {\bibfnamefont {J.}~\bibnamefont {Dalibard}}, \
  and\ \bibinfo {author} {\bibfnamefont {W.}~\bibnamefont {Zwerger}},\
  }\href@noop {} {\bibfield  {journal} {\bibinfo  {journal} {Rev. Mod. Phys.}\
  }\textbf {\bibinfo {volume} {80}},\ \bibinfo {pages} {885} (\bibinfo {year}
  {2008})}\BibitemShut {NoStop}%
\bibitem [{\citenamefont {Esteve}\ \emph {et~al.}(2006)\citenamefont {Esteve},
  \citenamefont {J.-B.Trebbia}, \citenamefont {Schumm}, \citenamefont {Aspect},
  \citenamefont {Westbrook},\ and\ \citenamefont {Bouchoule}}]{Esteve2006}%
  \BibitemOpen
  \bibfield  {author} {\bibinfo {author} {\bibfnamefont {J.}~\bibnamefont
  {Esteve}}, \bibinfo {author} {\bibnamefont {J.-B.Trebbia}}, \bibinfo {author}
  {\bibfnamefont {T.}~\bibnamefont {Schumm}}, \bibinfo {author} {\bibfnamefont
  {A.}~\bibnamefont {Aspect}}, \bibinfo {author} {\bibfnamefont {C.~I.}\
  \bibnamefont {Westbrook}}, \ and\ \bibinfo {author} {\bibfnamefont
  {I.}~\bibnamefont {Bouchoule}},\ }\href@noop {} {\bibfield  {journal}
  {\bibinfo  {journal} {Phys. Rev. Lett.}\ }\textbf {\bibinfo {volume} {96}},\
  \bibinfo {pages} {130403} (\bibinfo {year} {2006})}\BibitemShut {NoStop}%
\bibitem [{\citenamefont {Gemelke}\ \emph {et~al.}(2009)\citenamefont
  {Gemelke}, \citenamefont {Zhang}, \citenamefont {Hung},\ and\ \citenamefont
  {Chin}}]{Gemelke2009}%
  \BibitemOpen
  \bibfield  {author} {\bibinfo {author} {\bibfnamefont {N.}~\bibnamefont
  {Gemelke}}, \bibinfo {author} {\bibfnamefont {X.}~\bibnamefont {Zhang}},
  \bibinfo {author} {\bibfnamefont {C.-L.}\ \bibnamefont {Hung}}, \ and\
  \bibinfo {author} {\bibfnamefont {C.}~\bibnamefont {Chin}},\ }\href@noop {}
  {\bibfield  {journal} {\bibinfo  {journal} {Nature}\ }\textbf {\bibinfo
  {volume} {460}},\ \bibinfo {pages} {995} (\bibinfo {year}
  {2009})}\BibitemShut {NoStop}%
\bibitem [{\citenamefont {M\"uller}\ \emph {et~al.}(2010)\citenamefont
  {M\"uller}, \citenamefont {Zimmermann}, \citenamefont {Meineke},
  \citenamefont {Brantut}, \citenamefont {T},\ and\ \citenamefont
  {Moritz}}]{Muller2010}%
  \BibitemOpen
  \bibfield  {author} {\bibinfo {author} {\bibfnamefont {T.}~\bibnamefont
  {M\"uller}}, \bibinfo {author} {\bibfnamefont {B.}~\bibnamefont
  {Zimmermann}}, \bibinfo {author} {\bibfnamefont {J.}~\bibnamefont {Meineke}},
  \bibinfo {author} {\bibfnamefont {J.-P.}\ \bibnamefont {Brantut}}, \bibinfo
  {author} {\bibfnamefont {T.~E.}\ \bibnamefont {T}}, \ and\ \bibinfo {author}
  {\bibfnamefont {H.}~\bibnamefont {Moritz}},\ }\href@noop {} {\bibfield
  {journal} {\bibinfo  {journal} {Phys. Rev. Lett.}\ }\textbf {\bibinfo
  {volume} {105}},\ \bibinfo {pages} {040401} (\bibinfo {year}
  {2010})}\BibitemShut {NoStop}%
\bibitem [{\citenamefont {Bakr}\ \emph {et~al.}(2010)\citenamefont {Bakr},
  \citenamefont {Peng}, \citenamefont {Tai}, \citenamefont {Ma}, \citenamefont
  {Simon}, \citenamefont {Gillen}, \citenamefont {Folling}, \citenamefont
  {Pollet}, \citenamefont {M.},\ and\ \citenamefont {Greiner}}]{Bakr2010}%
  \BibitemOpen
  \bibfield  {author} {\bibinfo {author} {\bibfnamefont {W.~S.}\ \bibnamefont
  {Bakr}}, \bibinfo {author} {\bibfnamefont {A.}~\bibnamefont {Peng}}, \bibinfo
  {author} {\bibfnamefont {M.~E.}\ \bibnamefont {Tai}}, \bibinfo {author}
  {\bibfnamefont {R.}~\bibnamefont {Ma}}, \bibinfo {author} {\bibfnamefont
  {J.}~\bibnamefont {Simon}}, \bibinfo {author} {\bibfnamefont {J.~I.}\
  \bibnamefont {Gillen}}, \bibinfo {author} {\bibfnamefont {S.}~\bibnamefont
  {Folling}}, \bibinfo {author} {\bibfnamefont {L.}~\bibnamefont {Pollet}},
  \bibinfo {author} {\bibnamefont {M.}}, \ and\ \bibinfo {author} {\bibnamefont
  {Greiner}},\ }\href@noop {} {\bibfield  {journal} {\bibinfo  {journal}
  {Science}\ }\textbf {\bibinfo {volume} {329}},\ \bibinfo {pages} {547}
  (\bibinfo {year} {2010})}\BibitemShut {NoStop}%
\bibitem [{\citenamefont {Sherson}\ \emph {et~al.}(2010)\citenamefont
  {Sherson}, \citenamefont {Weitenberg}, \citenamefont {Endres}, \citenamefont
  {Cheneau}, \citenamefont {Bloch},\ and\ \citenamefont {Kuhr}}]{Sherson2010}%
  \BibitemOpen
  \bibfield  {author} {\bibinfo {author} {\bibfnamefont {J.~F.}\ \bibnamefont
  {Sherson}}, \bibinfo {author} {\bibfnamefont {C.}~\bibnamefont {Weitenberg}},
  \bibinfo {author} {\bibfnamefont {M.}~\bibnamefont {Endres}}, \bibinfo
  {author} {\bibfnamefont {M.}~\bibnamefont {Cheneau}}, \bibinfo {author}
  {\bibfnamefont {I.}~\bibnamefont {Bloch}}, \ and\ \bibinfo {author}
  {\bibfnamefont {S.}~\bibnamefont {Kuhr}},\ }\href@noop {} {\bibfield
  {journal} {\bibinfo  {journal} {Nature}\ }\textbf {\bibinfo {volume} {467}},\
  \bibinfo {pages} {68} (\bibinfo {year} {2010})}\BibitemShut {NoStop}%
\bibitem [{Note1()}]{Note1}%
  \BibitemOpen
  \bibinfo {note} {Nonmonotonic dependence on the strength of disorder has also
  been observed in the noise magnitude of equilibrium disordered systems~\cite
  {Cohen1992}}\BibitemShut {NoStop}%
\bibitem [{\citenamefont {Rahmani}(2013)}]{Rahmani2013}%
  \BibitemOpen
  \bibfield  {author} {\bibinfo {author} {\bibfnamefont {A.}~\bibnamefont
  {Rahmani}},\ }\href@noop {} {\bibfield  {journal} {\bibinfo  {journal} {Mod.
  Phys. Lett. B}\ }\textbf {\bibinfo {volume} {27}},\ \bibinfo {pages}
  {1330019} (\bibinfo {year} {2013})}\BibitemShut {NoStop}%
\bibitem [{\citenamefont {Altshuler}(1985)}]{Altshuler1985}%
  \BibitemOpen
  \bibfield  {author} {\bibinfo {author} {\bibfnamefont {B.~L.}\ \bibnamefont
  {Altshuler}},\ }\href@noop {} {\bibfield  {journal} {\bibinfo  {journal}
  {JETP Lett.}\ }\textbf {\bibinfo {volume} {41}},\ \bibinfo {pages} {468}
  (\bibinfo {year} {1985})}\BibitemShut {NoStop}%
\bibitem [{\citenamefont {Lee}\ and\ \citenamefont {Stone}(1985)}]{Lee1985}%
  \BibitemOpen
  \bibfield  {author} {\bibinfo {author} {\bibfnamefont {P.~A.}\ \bibnamefont
  {Lee}}\ and\ \bibinfo {author} {\bibfnamefont {A.~D.}\ \bibnamefont
  {Stone}},\ }\href@noop {} {\bibfield  {journal} {\bibinfo  {journal} {Phys.
  Rev. Lett.}\ }\textbf {\bibinfo {volume} {55}},\ \bibinfo {pages} {1622}
  (\bibinfo {year} {1985})}\BibitemShut {NoStop}%
\bibitem [{\citenamefont {Karpiuk}\ \emph {et~al.}(2012)\citenamefont
  {Karpiuk}, \citenamefont {Cherroret}, \citenamefont {Lee}, \citenamefont
  {Gr\'emaud}, \citenamefont {M\"uller},\ and\ \citenamefont
  {Miniatura}}]{Karpiuk2012}%
  \BibitemOpen
  \bibfield  {author} {\bibinfo {author} {\bibfnamefont {T.}~\bibnamefont
  {Karpiuk}}, \bibinfo {author} {\bibfnamefont {N.}~\bibnamefont {Cherroret}},
  \bibinfo {author} {\bibfnamefont {K.~L.}\ \bibnamefont {Lee}}, \bibinfo
  {author} {\bibfnamefont {B.}~\bibnamefont {Gr\'emaud}}, \bibinfo {author}
  {\bibfnamefont {C.~A.}\ \bibnamefont {M\"uller}}, \ and\ \bibinfo {author}
  {\bibfnamefont {C.}~\bibnamefont {Miniatura}},\ }\href@noop {} {\bibfield
  {journal} {\bibinfo  {journal} {Phys. Rev. Lett.}\ }\textbf {\bibinfo
  {volume} {109}},\ \bibinfo {pages} {190601} (\bibinfo {year}
  {2012})}\BibitemShut {NoStop}%
\bibitem [{\citenamefont {Micklitz}\ \emph {et~al.}(2014)\citenamefont
  {Micklitz}, \citenamefont {M\"uller},\ and\ \citenamefont
  {Altland}}]{Micklitz2014}%
  \BibitemOpen
  \bibfield  {author} {\bibinfo {author} {\bibfnamefont {T.}~\bibnamefont
  {Micklitz}}, \bibinfo {author} {\bibfnamefont {C.~A.}\ \bibnamefont
  {M\"uller}}, \ and\ \bibinfo {author} {\bibfnamefont {A.}~\bibnamefont
  {Altland}},\ }\href@noop {} {\bibfield  {journal} {\bibinfo  {journal} {Phys.
  Rev. Lett.}\ }\textbf {\bibinfo {volume} {112}},\ \bibinfo {pages} {110602}
  (\bibinfo {year} {2014})}\BibitemShut {NoStop}%
\bibitem [{\citenamefont {Ghosh}\ \emph {et~al.}(2014)\citenamefont {Ghosh},
  \citenamefont {Cherroret}, \citenamefont {Gr\'emaud}, \citenamefont
  {Miniatura},\ and\ \citenamefont {Delande}}]{Ghosh2014}%
  \BibitemOpen
  \bibfield  {author} {\bibinfo {author} {\bibfnamefont {S.}~\bibnamefont
  {Ghosh}}, \bibinfo {author} {\bibfnamefont {N.}~\bibnamefont {Cherroret}},
  \bibinfo {author} {\bibfnamefont {B.}~\bibnamefont {Gr\'emaud}}, \bibinfo
  {author} {\bibfnamefont {C.}~\bibnamefont {Miniatura}}, \ and\ \bibinfo
  {author} {\bibfnamefont {D.}~\bibnamefont {Delande}},\ }\href@noop {}
  {\bibfield  {journal} {\bibinfo  {journal} {Phys. Rev. A}\ }\textbf {\bibinfo
  {volume} {90}},\ \bibinfo {pages} {063602} (\bibinfo {year}
  {2014})}\BibitemShut {NoStop}%
\bibitem [{\citenamefont {Lee}\ \emph {et~al.}(2014)\citenamefont {Lee},
  \citenamefont {Gr\'emaud},\ and\ \citenamefont {Miniatura}}]{Lee2014}%
  \BibitemOpen
  \bibfield  {author} {\bibinfo {author} {\bibfnamefont {K.~L.}\ \bibnamefont
  {Lee}}, \bibinfo {author} {\bibfnamefont {B.}~\bibnamefont {Gr\'emaud}}, \
  and\ \bibinfo {author} {\bibfnamefont {C.}~\bibnamefont {Miniatura}},\
  }\href@noop {} {\bibfield  {journal} {\bibinfo  {journal} {Phys. Rev. A}\
  }\textbf {\bibinfo {volume} {90}},\ \bibinfo {pages} {043605} (\bibinfo
  {year} {2014})}\BibitemShut {NoStop}%
\bibitem [{\citenamefont {D'Alessio}\ and\ \citenamefont
  {Rahmani}(2013)}]{Dalessio2013}%
  \BibitemOpen
  \bibfield  {author} {\bibinfo {author} {\bibfnamefont {L.}~\bibnamefont
  {D'Alessio}}\ and\ \bibinfo {author} {\bibfnamefont {A.}~\bibnamefont
  {Rahmani}},\ }\href@noop {} {\bibfield  {journal} {\bibinfo  {journal} {Phys.
  Rev. B}\ }\textbf {\bibinfo {volume} {87}},\ \bibinfo {pages} {174301}
  (\bibinfo {year} {2013})}\BibitemShut {NoStop}%
\bibitem [{\citenamefont {Efetov}(1997)}]{Efetov1997}%
  \BibitemOpen
  \bibfield  {author} {\bibinfo {author} {\bibfnamefont {K.~B.}\ \bibnamefont
  {Efetov}},\ }\href@noop {} {\emph {\bibinfo {title} {Supersymmetry in
  Disorder and Chaos}}}\ (\bibinfo  {publisher} {Cambridge University Press},\
  \bibinfo {year} {1997})\BibitemShut {NoStop}%
\bibitem [{\citenamefont {Cuevas}\ and\ \citenamefont
  {Kravtsov}(2007)}]{Cuevas2007}%
  \BibitemOpen
  \bibfield  {author} {\bibinfo {author} {\bibfnamefont {E.}~\bibnamefont
  {Cuevas}}\ and\ \bibinfo {author} {\bibfnamefont {V.~E.}\ \bibnamefont
  {Kravtsov}},\ }\href@noop {} {\bibfield  {journal} {\bibinfo  {journal}
  {Phys. Rev. B}\ }\textbf {\bibinfo {volume} {76}},\ \bibinfo {pages} {235119}
  (\bibinfo {year} {2007})}\BibitemShut {NoStop}%
\bibitem [{\citenamefont {Sanchez-Palencia}\ and\ \citenamefont
  {Lewenstein}(2010)}]{Sanchez2010}%
  \BibitemOpen
  \bibfield  {author} {\bibinfo {author} {\bibfnamefont {L.}~\bibnamefont
  {Sanchez-Palencia}}\ and\ \bibinfo {author} {\bibfnamefont {M.}~\bibnamefont
  {Lewenstein}},\ }\href@noop {} {\bibfield  {journal} {\bibinfo  {journal}
  {Nat. Phys.}\ }\textbf {\bibinfo {volume} {6}},\ \bibinfo {pages} {87}
  (\bibinfo {year} {2010})}\BibitemShut {NoStop}%
\bibitem [{\citenamefont {Shapiro}(2012)}]{Shapiro2012}%
  \BibitemOpen
  \bibfield  {author} {\bibinfo {author} {\bibfnamefont {B.}~\bibnamefont
  {Shapiro}},\ }\href@noop {} {\bibfield  {journal} {\bibinfo  {journal} {J. of
  Phys. A: Mathematical and Theoretical}\ }\textbf {\bibinfo {volume} {45}},\
  \bibinfo {pages} {143001} (\bibinfo {year} {2012})}\BibitemShut {NoStop}%
\bibitem [{\citenamefont {Billy}\ \emph {et~al.}(2008)\citenamefont {Billy},
  \citenamefont {Josse}, \citenamefont {Zuo}, \citenamefont {Bernard},
  \citenamefont {Hambrecht}, \citenamefont {Lugan}, \citenamefont {Cl\'ement},
  \citenamefont {Sanchez-Palencia}, \citenamefont {Bouyer},\ and\ \citenamefont
  {Aspect}}]{Billy2008}%
  \BibitemOpen
  \bibfield  {author} {\bibinfo {author} {\bibfnamefont {J.}~\bibnamefont
  {Billy}}, \bibinfo {author} {\bibfnamefont {V.}~\bibnamefont {Josse}},
  \bibinfo {author} {\bibfnamefont {Z.}~\bibnamefont {Zuo}}, \bibinfo {author}
  {\bibfnamefont {A.}~\bibnamefont {Bernard}}, \bibinfo {author} {\bibfnamefont
  {B.}~\bibnamefont {Hambrecht}}, \bibinfo {author} {\bibfnamefont
  {P.}~\bibnamefont {Lugan}}, \bibinfo {author} {\bibfnamefont
  {D.}~\bibnamefont {Cl\'ement}}, \bibinfo {author} {\bibfnamefont
  {L.}~\bibnamefont {Sanchez-Palencia}}, \bibinfo {author} {\bibfnamefont
  {P.}~\bibnamefont {Bouyer}}, \ and\ \bibinfo {author} {\bibfnamefont
  {A.}~\bibnamefont {Aspect}},\ }\href@noop {} {\bibfield  {journal} {\bibinfo
  {journal} {Nature}\ }\textbf {\bibinfo {volume} {453}},\ \bibinfo {pages}
  {891} (\bibinfo {year} {2008})}\BibitemShut {NoStop}%
\bibitem [{\citenamefont {Roati}\ \emph {et~al.}(2008)\citenamefont {Roati},
  \citenamefont {D’Errico}, \citenamefont {Fallani}, \citenamefont {Fattori},
  \citenamefont {Fort}, \citenamefont {Zaccanti}, \citenamefont {Modugno},
  \citenamefont {Modugno},\ and\ \citenamefont {Inguscio}}]{Roati2008}%
  \BibitemOpen
  \bibfield  {author} {\bibinfo {author} {\bibfnamefont {G.}~\bibnamefont
  {Roati}}, \bibinfo {author} {\bibfnamefont {C.}~\bibnamefont {D’Errico}},
  \bibinfo {author} {\bibfnamefont {L.}~\bibnamefont {Fallani}}, \bibinfo
  {author} {\bibfnamefont {M.}~\bibnamefont {Fattori}}, \bibinfo {author}
  {\bibfnamefont {C.}~\bibnamefont {Fort}}, \bibinfo {author} {\bibfnamefont
  {M.}~\bibnamefont {Zaccanti}}, \bibinfo {author} {\bibfnamefont
  {G.}~\bibnamefont {Modugno}}, \bibinfo {author} {\bibfnamefont
  {M.}~\bibnamefont {Modugno}}, \ and\ \bibinfo {author} {\bibfnamefont
  {M.}~\bibnamefont {Inguscio}},\ }\href@noop {} {\bibfield  {journal}
  {\bibinfo  {journal} {Nature}\ }\textbf {\bibinfo {volume} {453}},\ \bibinfo
  {pages} {891} (\bibinfo {year} {2008})}\BibitemShut {NoStop}%
\bibitem [{\citenamefont {Kondov}\ \emph {et~al.}(2011)\citenamefont {Kondov},
  \citenamefont {McGehee}, \citenamefont {Zirbel},\ and\ \citenamefont
  {DeMarco}}]{Kondov2011}%
  \BibitemOpen
  \bibfield  {author} {\bibinfo {author} {\bibfnamefont {S.~S.}\ \bibnamefont
  {Kondov}}, \bibinfo {author} {\bibfnamefont {W.~R.}\ \bibnamefont {McGehee}},
  \bibinfo {author} {\bibfnamefont {J.~J.}\ \bibnamefont {Zirbel}}, \ and\
  \bibinfo {author} {\bibfnamefont {B.}~\bibnamefont {DeMarco}},\ }\href@noop
  {} {\bibfield  {journal} {\bibinfo  {journal} {Science}\ }\textbf {\bibinfo
  {volume} {334}},\ \bibinfo {pages} {66} (\bibinfo {year} {2011})}\BibitemShut
  {NoStop}%
\bibitem [{\citenamefont {McGehee}\ \emph {et~al.}(2013)\citenamefont
  {McGehee}, \citenamefont {Kondov}, \citenamefont {Xu}, \citenamefont
  {Zirbel},\ and\ \citenamefont {DeMarco}}]{McGehee2013}%
  \BibitemOpen
  \bibfield  {author} {\bibinfo {author} {\bibfnamefont {W.~R.}\ \bibnamefont
  {McGehee}}, \bibinfo {author} {\bibfnamefont {S.~S.}\ \bibnamefont {Kondov}},
  \bibinfo {author} {\bibfnamefont {W.}~\bibnamefont {Xu}}, \bibinfo {author}
  {\bibfnamefont {J.~J.}\ \bibnamefont {Zirbel}}, \ and\ \bibinfo {author}
  {\bibfnamefont {B.}~\bibnamefont {DeMarco}},\ }\href@noop {} {\bibfield
  {journal} {\bibinfo  {journal} {Phys. Rev. Lett.}\ }\textbf {\bibinfo
  {volume} {111}},\ \bibinfo {pages} {145303} (\bibinfo {year}
  {2013})}\BibitemShut {NoStop}%
\bibitem [{\citenamefont {Cohen}\ \emph {et~al.}(1992)\citenamefont {Cohen},
  \citenamefont {Ovadyahu},\ and\ \citenamefont {Rokni}}]{Cohen1992}%
  \BibitemOpen
  \bibfield  {author} {\bibinfo {author} {\bibfnamefont {O.}~\bibnamefont
  {Cohen}}, \bibinfo {author} {\bibfnamefont {Z.}~\bibnamefont {Ovadyahu}}, \
  and\ \bibinfo {author} {\bibfnamefont {M.}~\bibnamefont {Rokni}},\
  }\href@noop {} {\bibfield  {journal} {\bibinfo  {journal} {Phys. Rev. Lett.}\
  }\textbf {\bibinfo {volume} {69}},\ \bibinfo {pages} {3555} (\bibinfo {year}
  {1992})}\BibitemShut {NoStop}%
\end{thebibliography}%

\end{document}